\documentclass[fleqn,usenatbib]{mnras}

\usepackage{newtxtext,newtxmath}
\usepackage[T1]{fontenc}

\DeclareRobustCommand{\VAN}[3]{#2}
\let\VANthebibliography\thebibliography
\def\thebibliography{\DeclareRobustCommand{\VAN}[3]{##3}\VANthebibliography}

\usepackage{graphicx}	
\usepackage{amsmath}	
\usepackage{gensymb}
\usepackage{xspace}

\graphicspath{{fig/}}
\newcommand{\pa}{\partial}
\newcommand{\Rsun}{$R_\odot$\xspace}

\title[Theory of Near--Surface Shear Layer]{A Theoretical Model of the Near--Surface Shear Layer of the Sun}

\author[B. K. Jha et al.]{
Bibhuti Kumar Jha,$^{1,2,3}$\thanks{E-mail: maitraibibhu@gmail.com}
Arnab Rai Choudhuri,$^{4}$\thanks{E-mail: arnab@iisc.ac.in}
\\
$^{1}$Indian Institute of Astrophysics, Bangalore 560034, Karnataka,  India\\
$^{2}$Aryabhatta Research Institute of Observational Sciences, Nainital 263001, Uttarakhand, India\\
$^{3}$Pondicherry University, Chinna Kalapet, Kalapet, Puducherry 605014, India\\
$^{4}$Department of Physics, Indian Institute of Science, Bangalore 560012, Karnataka, India\\
}

\date{Accepted XXX. Received YYY; in original form ZZZ}

\pubyear{2021}

\begin{document}
\label{firstpage}
\pagerange{\pageref{firstpage}--\pageref{lastpage}}
\maketitle

\begin{abstract}
The Sun has a Near-Surface Shear Layer (NSSL), within which the angular velocity decreases rapidly with radius. We provide an explanation of this layer based on the thermal wind balance equation.  Since convective motions are not affected by solar rotation in the top layer of the convection zone, we argue that the temperature falls at the same rate at all latitudes in this layer.  This makes the thermal wind term very large in this layer and the centrifugal term has also to become very large to balance it, giving rise to the NSSL.  From the values of differential rotation $\Omega (r<r_c, \theta)$ at radii less than a radius $r_c$, we can calculate the temperature difference $\Delta T (r, \theta)$ with respect to the standard solar model at different points of the convection zone, by making use of the thermal wind balance equation.  Then we again use this equation in the top layer to calculate $\Omega (r>r_c, \theta)$ there from $\Delta T (r, \theta)$. We carry on this exercise using both an analytical expression of the differential rotation and the actual data provided by helioseismology. We find that our theoretical results of the NSSL match the observational data reasonably well for $r_c \approx 0.96$\Rsun, giving an estimate of the radius till which the convective motions are affected by the solar rotation.
\end{abstract}




\section{Introduction}
\label{sec:intro}
\begin{figure}
	\includegraphics[width=\columnwidth]{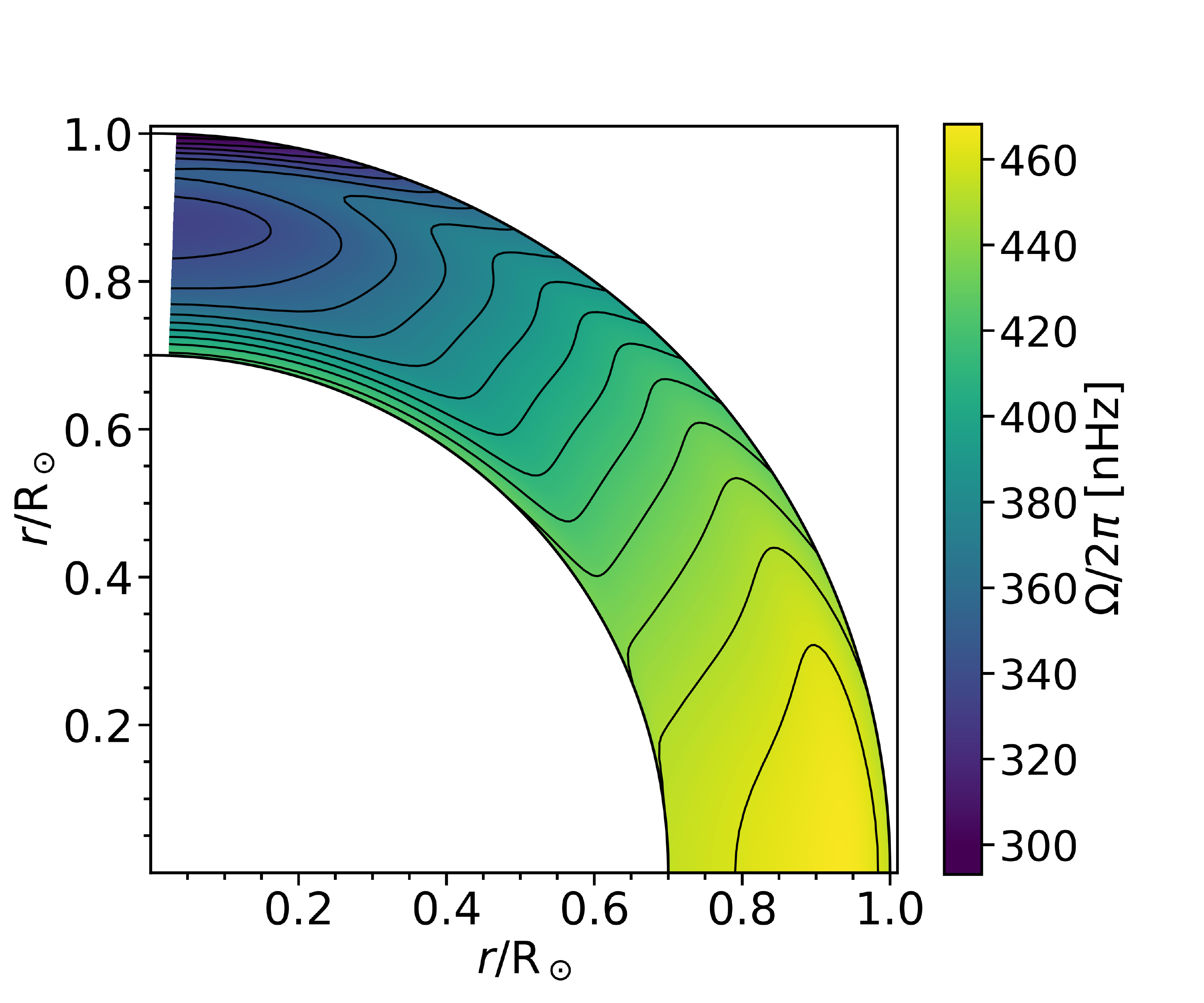}
    \caption{Rotation profile on the basis of helioseismology data. The contours correspond
    to values of $\Omega$ at the interval of 10 nHz,
    the extreme contours inside the main body
    of the convection zone being for the values 340 nHz
    and 460 nHz (apart from a few lower value contours
    near the surface at the polar region, where
    helioseismic inversions are not trustworthy).}
    \label{fig:omega_helios}
\end{figure}

One of the intriguing features in the differential rotation map of the Sun, as seen, for example, in Figure~1 of \citet{howe_2009} or in Figure~26 of \citet{Basu2016}, is the existence of the near-surface shear layer (NSSL). This is a layer near the solar surface at the top of the convection zone, within which the angular velocity decreases sharply with increasing solar radius.  The first indication of the existence of such a layer came more than half a century ago, when it was noted that the rotation rate of the solar surface measured from the Doppler shifts of photospheric spectral lines was about 5\% lower than the rotation rate inferred from the positions of sunspots on the solar surface \citep{Howard_Harvey1970}. While the depth at which sunspots are anchored remains unclear and probably changes with the age of a sunspot group \citep{Longcope_Choudhuri2002}, the rotation rate inferred from the sunspots was assumed to correspond to a layer underneath the solar surface, implying that the angular velocity was higher in that layer. When helioseismology mapped the internal differential rotation of the Sun, the existence of this layer was fully established.  Figure~\ref{fig:omega_helios} shows the differential rotation map
of the Sun obtained by helioseismology (with contours of constant angular velocity) which we shall use later in our calculations in Section~\ref{sec:helio}.  The contours of constant angular velocity, which are nearly radial within a large part of the body of the solar convection zone, bend towards the equator within a layer of thickness of order $\approx 0.05$\Rsun at the top of the convection zone \citep{Schou1998, Howe2005}.

What causes this NSSL is still not properly understood.  The first attempts to explain it \citep{Foukal1975, Gilman1979} were based on the idea that convection in the upper layers of the convection zone mixes angular momentum in such a manner that the angular momentum per unit mass tends to become constant in these layers, leading to a decrease of the angular velocity with radius.  Once the differential rotation of the Sun was properly mapped and no evidence was found for the constancy of the specific
angular momentum within the convection zone, it was realized that this
could not be the appropriate explanation. Within the last few years, there have been attempts to explain the NSSL on the basis of numerical simulations of the solar convection \citep{guerrero_2013,hotta_2015, Matilsky2019}. It has been argued by \citet{hotta_2015} that the Reynolds stresses play an important role in creating the NSSL, whereas \citet{Matilsky2019} suggested that the steep decrease in density in the top layers of the convection zone is crucial in giving rise to the NSSL. \citet{Choudhuri2021a} has recently proposed a possible alternative theoretical explanation of the NSSL based on order-of-magnitude estimates. The aim of the present paper is to substantiate the ideas proposed by \citet{Choudhuri2021a} through detailed calculations.

The theories of the two large-scale flow patterns within the convection zone of the Sun---the differential rotation and the meridional circulation---are intimately connected with each other \citep{kitchatinov_2013, Choudhuri2021b}.  The idea we wish to develop follows from the central equation in the theory of the meridional circulation: the thermal wind balance equation. Since the nature of the Coriolis force arising out of the solar rotation varies with latitude, the effect of this force on the convection is expected to vary with latitude \citep{durney_1971, belvedere_1976}. Since the Coriolis force provides the least hindrance to convective heat transport in the polar regions, the poles of the Sun are expected to be slightly hotter than the equator \citep{kitchatinov_1995}. There are some observational indications that this may indeed be the case  \citep{kuhn_1988, rast_2008}. Hotter poles would tend to drive what is called a thermal wind, i.e.\ a meridional circulation which would be equatorward at the solar surface. Since the observed meridional circulation is the opposite of that, we must have another effect which overpowers this and drives the meridional circulation in the poleward direction at the surface as observed. It is easy to show that the centrifugal force arising out of the observed differential rotation of the Sun can do this job: see Figure~8 and the accompanying text in \citet{Choudhuri2021b}. The term corresponding to the dissipation of the meridional circulation is found to be negligible compared to the driving terms within the bulk of the convention zone, as pointed out in the Appendix of \citet{Choudhuri2021b}.  As a result, we expect the two driving terms of the meridional circulation---the thermal wind term and the centrifugal term---to be comparable within the main body of the solar convection zone.  This is often referred to as the thermal wind balance condition.

It is generally believed that the thermal wind balance condition holds within the body of the convection zone
\citep{kitchatinov_2013, karak2014b}, although different authors may not completely agree as to the extent to which 
it holds  \citep{Brun2010}.  
There is, however, not much agreement among different authors whether the thermal wind condition should
hold even within the top upper layer of the solar convection zone. A widely held view is that this upper
layer is a kind of boundary layer within which the dissipation term or Reynolds stresses become important, giving rise to a
violation of the thermal wind balance condition.  It is argued that the NSSL arises in some manner out of this
violation.  A completely opposite argument is given in the earlier paper by \citet{Choudhuri2021a} and in the
present paper.  We point out that the thermal wind term becomes very large in the top layer of the
solar convection zone due to a combination of two factors: (i) the temperature falls sharply as we move outward
through this layer, and (ii) the pole-equator temperature difference does not vary with depth in this
layer because of the reduced effect of the Coriolis force on convection in this layer, as explained
in the next section (Section~\ref{sec:method}). The dissipation term is much smaller than the
thermal wind term even within the main body of the convection zone, as the order of magnitude
estimate in the Appendix of \citet{Choudhuri2021a} suggests. If the thermal wind term becomes
even much larger in the upper layers of the convection zone, then it appears unlikely to us that this large thermal wind term
can be balanced by the dissipation term.  The only possibility is that the centrifugal term also
has to become very large in the top layer to balance the thermal wind term.  This dictates that the
top layer has to be a region within which the angular velocity undergoes a large variation.  We show
through quantitative calculations that the structure of the NSSL calculated theoretically on the basis
of our ideas agrees with the observational data remarkably well.

We explain our basic methodology in the next section (Section~\ref{sec:method}).  After that Section~\ref{sec:empirical} is devoted to the applying
our methodology to an analytical expression of the differential rotation in the interior of the solar
convection zone.  Then the actual data of differential rotation obtained by helioseismology are applied to
calculate the structure of the NSSL in Section~\ref{sec:helio}. Finally, our conclusions are summarized in Section~\ref{sec:conclusion}. 

\section{Basic Methodology}
\label{sec:method}

The equation for thermal wind balance is
\begin{equation}
r \sin{\theta} \frac{\pa}{\pa z} \Omega^2=\frac{1}{r}\frac{g}{\gamma C_V} \frac{\pa S}{\pa \theta},
\label{eq1}
\end{equation}
where $\Omega$ is the angular velocity, $z$ is the distance from the equatorial plane
measured upward, $g$ is the acceleration due to gravity of the Sun at the point
under consideration and $\gamma$ is the adiabatic index, while $S$ and $C_V$ are respectively the entropy and the specific heat of the gas per
unit mass. See \citet{Choudhuri2021b} for the derivation and a discussion of this equation.
We point out that the Sun is expected to have a small oblateness
due to its rotation.  Although we are not aware of any direct
measurements of this oblateness, we discuss in Appendix~\ref{appendix}
how it can be estimated from theoretical considerations. 
We show that this oblateness should be
less than 0.0005\% and we also argue in Appendix~\ref{appendix} that
the observed solar surface would be an isochoric surface in
the absence of the thermal wind.  To be able to compare with
observational data of the solar surface, we should consider
temperature variations over isochoric surfaces rather than over
surfaces with constant $r$.  This actually simplifies things
because the entropy and temperature differentials on an isochoric
surface are related by
\begin{equation}
\Delta S = C_V \frac{\Delta T}{T},
\label{eq2}
\end{equation}
We also point out in Appendix~\ref{appendix} that 
Equation~\ref{eq1} and Equation~\ref{eq2} give the following equation for the temperature
differences on isochoric surfaces
\begin{equation}
r^2 \sin \theta \frac{\pa}{\pa z} \Omega^2 = \frac{g}{\gamma T} \left( \frac{d}{d \theta} \Delta T \right)_{\rm isochore}.
\label{eq3}
\end{equation}
This is the main equation on which the analysis of the present paper is based. Since the oblateness of isochoic surfaces
in the Sun is so small, they can be regarded as spherical surfaces
for most practical purposes.  However, from a conceptual point
of view, it is important to remember that our central equation (Equation~\ref{eq3})
refers to temperature variations over isochoic surfaces and
that is what gives the possibility of comparing our theoretical
results with observational data of the solar surface.

In our discussions, we sometimes will have to deal with
situations like the following. We may know the distribution of
$\Omega (r, \theta)$ in some region. Suppose we also know 
the temperature on some axis $\theta = \theta_0$, which means
that we would know the value of the temperature at one point on
an isochore.  We want to find the temperature at other points
on the isochore.  Although an isochore is not a surface of constant
$r$ at a conceptual level, we can identify an isochore by a value
of $r$ due to its very small oblateness.  According to Equation~\ref{eq3}, the
temperature difference between the points $(r, \theta)$ and $(r,\theta_0)$
on an isochore is given by  
\begin{equation}
\Delta T_{\theta_0} (r, \theta) = \frac{r^2 \gamma}{g} 
\int_{\theta_0}^{\theta} d\theta \; T
\sin \theta \frac{\pa}{\pa z} \Omega^2.
\label{eq34}
\end{equation}
Because of the small oblateness of the isochore, we can think of
Equation~\ref{eq34} as an integration along a circle of constant $r$ and we shall
later represent $\Delta T_{\theta_0} (r, \theta)$ graphically as if
an isochore is a circule of constant $r$. However, at a conceptual
level, we should keep in mind that $\Delta T_{\theta_0} (r, \theta)$
gives the temperature variation over an isochore rather than over a
circle of constant $r$.

As stressed earlier by \citet{Choudhuri2021b, Choudhuri2021a},
whether the Coriolis force due to the Sun's rotation has any effect on
the convection cells depends on whether the convective turnover time is
comparable to the rotation period or not.  Numerical simulations suggest 
that convection in the deeper layers of the convection zone involves large
convection cells with long turnover times and are affected by the Coriolis force:
see Figure~1 in \citet{Brown2010} or Figure~3 in \citet{gastine_2014}. As a result,
heat transfer depends on latitude.  However, this is not the case near
the top of the convection zone, where the convection cells (the granules) are
much smaller in size and have turnover times as short as a few minutes.
The extent to which the temperature gradient $d T/ dr$ differs from the
adiabatic gradient depends on the mixing length (see, for exmaple, \citet{Kippenhahn1990}, Section~7).  In the top of the 
convection zone which is not affected by rotation, the mixing length is
independent of latitute and we expect $d T/ dr$ also to be
independent of latitude \citep{Choudhuri2021a}.  Although there must be a gradual transition
from deeper layers within which heat transport depends on the latitude to the
top layer within which this is not the case, we assume for simplicity that the
transition takes place at radius $r = r_c$ above which we
have $d T/d r$ independent of latitude.  We shall work out our model
by assuming different values of $r_c$ in the range 0.92\,--\,0.98~\Rsun.

In the simplest kind of a spherically symmetric model of the Sun, the
temperature $T$ would be function of $r$ alone and would be independent of
$\theta$.  If the heat transport depends on latitude, then that would introduce
a small variation of $T$ with $\theta$. We can write
\begin{equation}
T(r, \theta) = T (r, 0) + \Delta T (r, \theta).
\label{eq4}
\end{equation}
If $d T/ dr$ is independent of latitude in the layer above $r > r_c$, then we have
\begin{equation}
\frac{dT (r,\theta)}{dr} = \frac{dT (r,0)}{dr} \nonumber
\end{equation}
so that it follows from Equation~\ref{eq4} that
\begin{equation}
\frac{d}{dr} \Delta T (r, \theta)= 0 \nonumber
\end{equation}
in this top layer.  So we can write
\begin{equation}
\Delta T (r > r_c, \theta) = \Delta T (r_c, \theta).
\label{eq5}
\end{equation}

In principle, it would be possible to determine $T (r, \theta)$ throughout the convection zone if we have a theory of how convective heat transport varies with latitude due to the effect of the Coriolis force.  Since our understanding of this complex problem is limited, we can proceed in a different manner. Since there is general agreement that the thermal wind balance condition holds within the deeper layers of the convection zone, we assume Equation~\ref{eq3} to hold below the radius $r = r_c$.  If we know the angular velocity $\Omega (r, \theta)$ in this region, then it is straightforward to evaluate the left hand side of Equation~\ref{eq3}.  Once we have the value of the left hand side, we can carry on integrations along different isochores in accordance with Equation~\ref{eq34} to determine $\Delta T (r, \theta)$ at all points within the convection zone below $ r = r_c$. Once we have the value of $\Delta T (r, \theta)$ at the radius $r = r_c$, we readily have the value of $\Delta T (r, \theta)$ at all points above this surface by using Equation~\ref{eq5}.  In other words, we can obtain $\Delta T (r, \theta)$ throughout the convection zone from the values of $\Omega (r, \theta)$ in the deeper layers below $r = r_c$, where Equation~\ref{eq3} is expected to hold.Because of the miniscule oblateness of the isochores, we are regarding the isochores as surfaces of constant $r$. However, it should be kept in mind that $\Delta T (r, \theta)$ we 
would be obtaining in this method refers to variations of temperature
over isochoric surfaces.  Comparing Equation~\ref{eq34} with Equation~\ref{eq5}, it should be clear
that $\Delta T (r, \theta)$ is nothing but $\Delta T_{\theta_0} (r, \theta)$ with $\theta_0 = 0$.

As we already pointed out, there is a lack of consensus whether the thermal wind balance equation holds in the top layer of the convection zone.  It follows from Equation~\ref{eq5} that $(\pa/\pa \theta) \Delta T (r,\theta)$ does not vary with $r$ above the radial surface $r = r_c$.  On the other hand, the temperature scale height becomes very small in this top layer and the temperature falls by orders of magnitude as we move to the solar surface from $r = r_c$. As a result, the thermal wind term represented by the right hand side of Equation~\ref{eq3} in which $T$ appears in the denominator becomes very large.  We do not think that this term can be balanced by the dissipation term.  We suggest that the thermal wind balance must hold even in this top layer and the centrifugal term represented by left hand side of Equation~\ref{eq3} has to become very large to balance the thermal wind term, implying a strong variation of $\Omega^2$ along $z$. As we have the values of $\Delta T$ above $r = r_c$, we can evaluate the right hand side of Equation~\ref{eq3} easily.  Then we can use Equation~\ref{eq3} to determine how $\Omega^2$ varies within this top layer.

In a nutshell, our methodology is as follows.  We start by assuming a value of $r = r_c$ below which convective heat transport is affected by the Coriolis force and above which this is not the case.  To begin with, we need the value of $\Omega (r, \theta)$ below $r_c$, from which we can calculate the left hand side of Equation~\ref{eq3} and eventually obtain $\Delta T (r, \theta)$ throughout the solar convection zone, obtaining $\Delta T (r, \theta)$ above $r_c$ by using Equation~\ref{eq5}.  Once we have $\Delta T (r, \theta)$ above $r_c$, the right hand side of Equation~\ref{eq3} can be evaluated, which enables us to find out $\Omega (r, \theta)$ above $r_c$ from Equation~\ref{eq3}. Although we use Equation~\ref{eq3} for all our calculations, we proceed differently below and above $r = r_c$.  Below $r_c$ we calculate $\Delta T (r, \theta)$ from $\Omega (r, \theta)$ beginning with the left hand side of Equation~\ref{eq3}, whereas above $r_c$ we calculate $\Omega (r, \theta)$ from $\Delta T (r, \theta)$ beginning with the right hand side of Equation~\ref{eq3}.

\begin{figure}
	\includegraphics[width=\columnwidth]{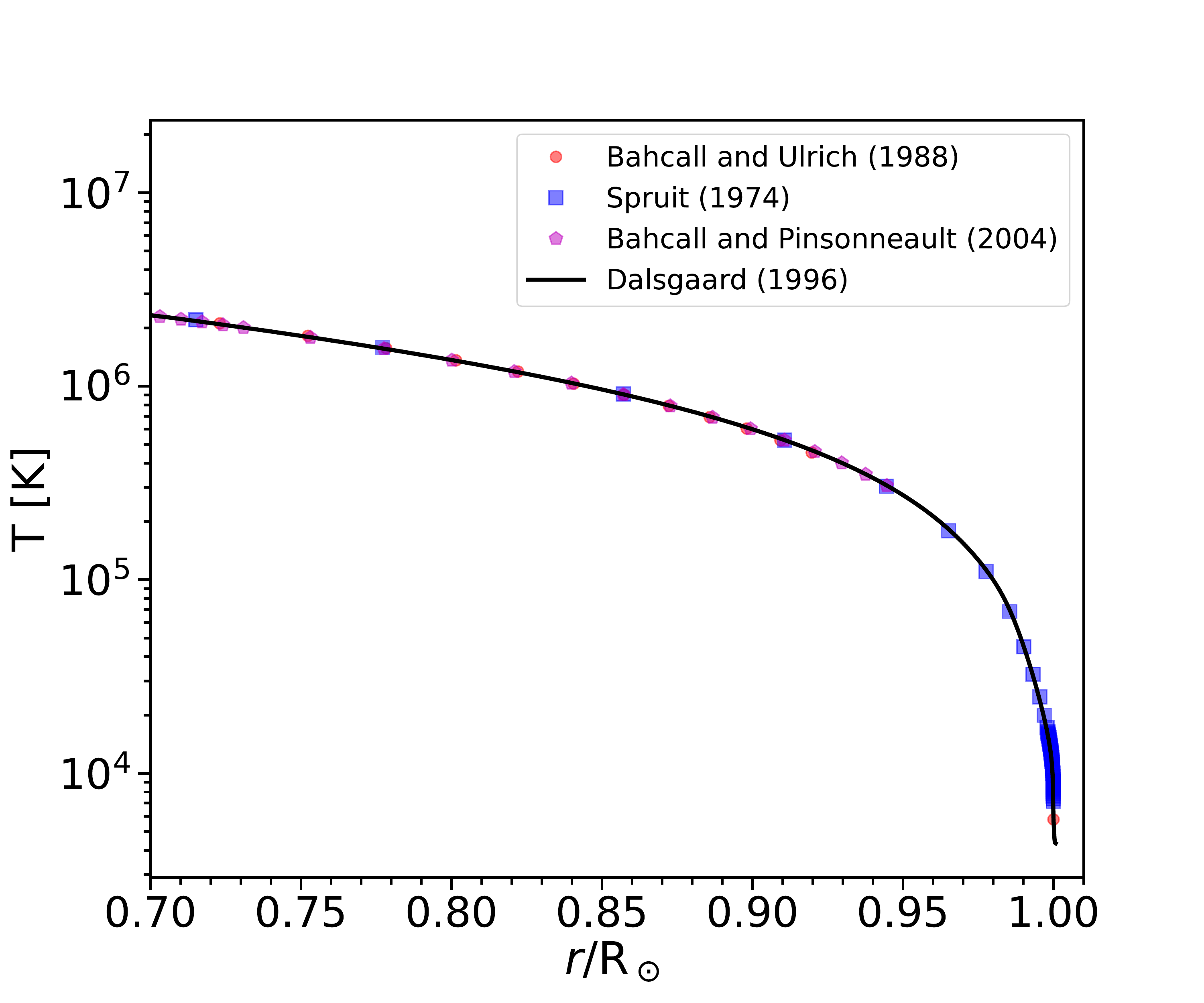}
    \caption{Variation of temperature with $r$ in different convection zone models. The curve corresponds
    to Model S used in our calculations.
    }
    \label{fig:t_prof}
\end{figure}

\begin{figure}
	\includegraphics[width=\columnwidth]{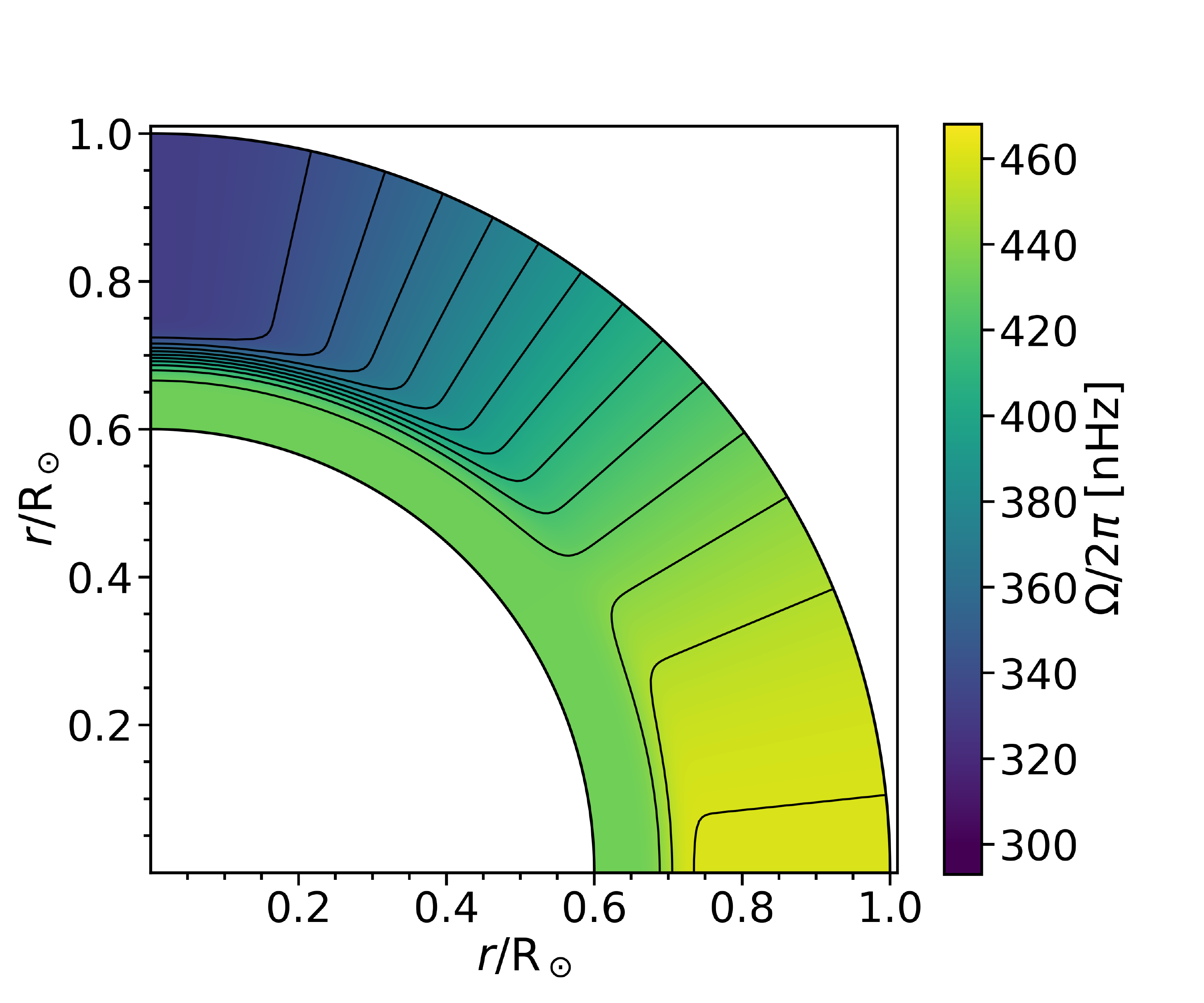}
    \caption{Rotation profile calculated by using the analytical expression (Equation~\ref{eq6}).
    The contours correspond
    to values of $\Omega$ at the interval of 10 nHz,
    the extreme contours being for the values 340 nHz
    and 460 nHz.}
    \label{fig:rot_prof_emp}
\end{figure}

The earlier paper by \citet{Choudhuri2021a} presented some order-of-magnitude estimates based on the methodology outlined above.  Now we present a detailed analysis. It order to carry on this analysis, we need the values of the temperature as a function of $r$, which we can take to be $T(r, 0)$, i.e.\ temperature values on the polar axis where the effect of the Coriolis force is minimal.  Several models of the convection zone exist in the literature  \citep{spruit_1974, bahcall_1988, Dalsgaard1996, Bahcall2004}. We use what has been referred to as Model~S by \citet{Dalsgaard1996} to obtain $T$ at different values of $r$.
Actually, all the models of the convection zone give very similar $T(r)$, as can be seen in Figure~\ref{fig:t_prof}.
We also note the sharp fall of the temperature in the outer layers of the convection zone, 
which is of crucial importance in our theory. The thermal wind term appearing in Equation~\ref{eq3} has $T$ in the denominator and becomes very large in the uppermost layers of the convection zone.  The value of the adiabatic
index $\gamma$ is taken to be 5/3 in all our calculations.  We point out that, in Model S, the value of $\gamma$ is very close to 5/3 throughout the convection zone except in the top layer 0.97\Rsun--\Rsun, where it becomes somewhat less due to the variations in the level of hydrogen ionization. We also need the values of $\Omega (r, \theta)$ below $r = r_c$ to start our calculations.  Calculations based on an analytical expression of  $\Omega (r, \theta)$ which fits helioseismology observtions reasonably well are presented in Section~\ref{sec:empirical}.  Then Section~\ref{sec:helio} will present calculations done with the actual helioseismology data of differential rotation $\Omega (r, \theta)$ used below $r_c$.  Calculations of both Section~\ref{sec:empirical} and Section~\ref{sec:helio} give the NSSL matching the observational data quite closely for appropriate values of $r_c$.  We point out that the input data of $\Omega (r, \theta)$ used in both these
sections (given by Equation~\ref{eq6} and from helioseismology respectively) give nearly radial contours till $r_c$ without much sign of the NSSL below $r_c$.  In fact, the analytical expression of $\Omega (r, \theta)$ that we use does not incorporate the NSSL at all and gives radial contours till the solar surface.  As we get the NSSL even in this case, we can clearly rule out the possibility that there might have been some indication about the existence of the NSSL in the input data which percolated through the calculations to give the NSSL at the end.  There is no doubt that the NSSL arises primarily out of the requirement that the centrifugal term has to match the thermal wind term which has become very large in the top layers of the solar convection zone.

It may be mentioned that \citet{Matilsky2020} calculated the temperature difference $\Delta T$ which one would get from the solar differential rotation by assuming the thermal wind balance Equation~\ref{eq3} to be valid till the top of the convection zone and plotted it in Figure~13 of their paper.  However, they did not discuss any physical significance of this. Some related issues are also discussed in a recent paper by \citet{Vasil2020}.

\begin{figure}
	\includegraphics[width=\columnwidth]{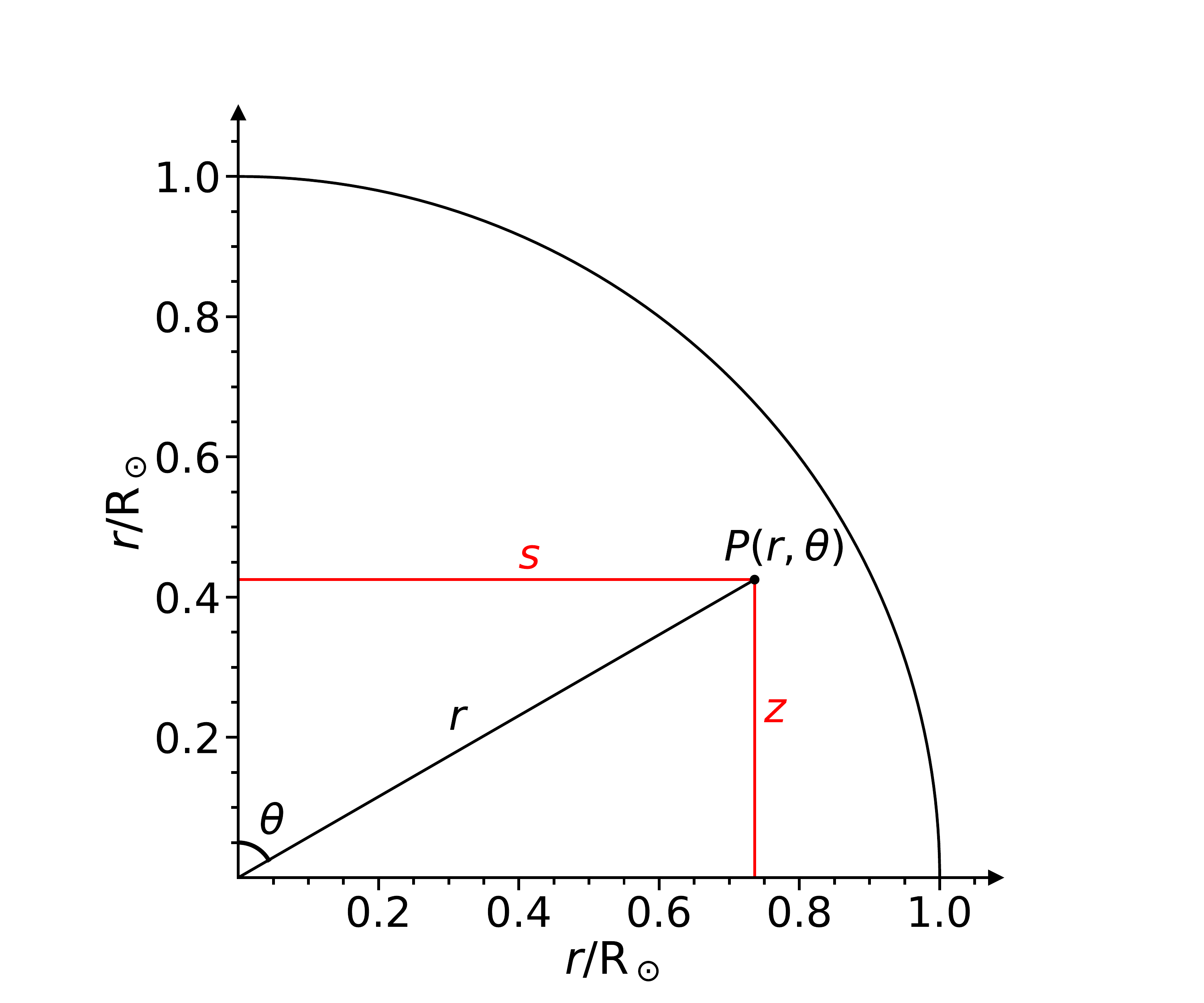}
    \caption{Relation between spherical co-ordinate system $(r, \theta)$ and the other coordinates $(s, z)$ we use.}
    \label{fig:context}
\end{figure}

\begin{figure}
	\includegraphics[width=\columnwidth]{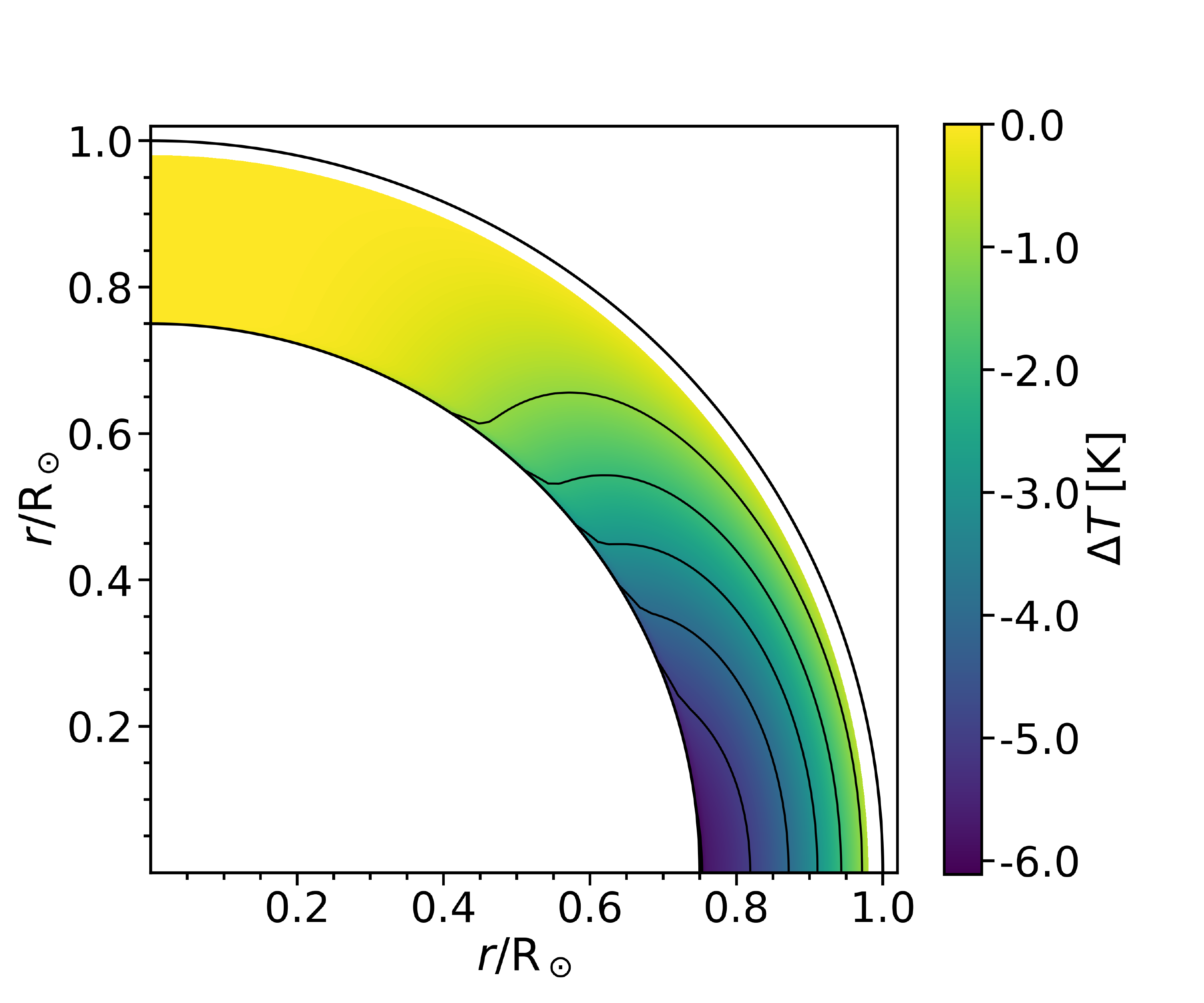}
    \caption{Profile of $\Delta T (r,\theta)$ in the solar convection zone
    below $r = 0.98$~\Rsun, which would follow from the analytical expression (Equation~\ref{eq6})
    on taking $r_c = 0.98$~\Rsun. Contours represent the constant values of $\Delta T (r,\theta)$.}
    \label{fig:delta_T_emp}
\end{figure}

\begin{figure}
	\includegraphics[width=\columnwidth]{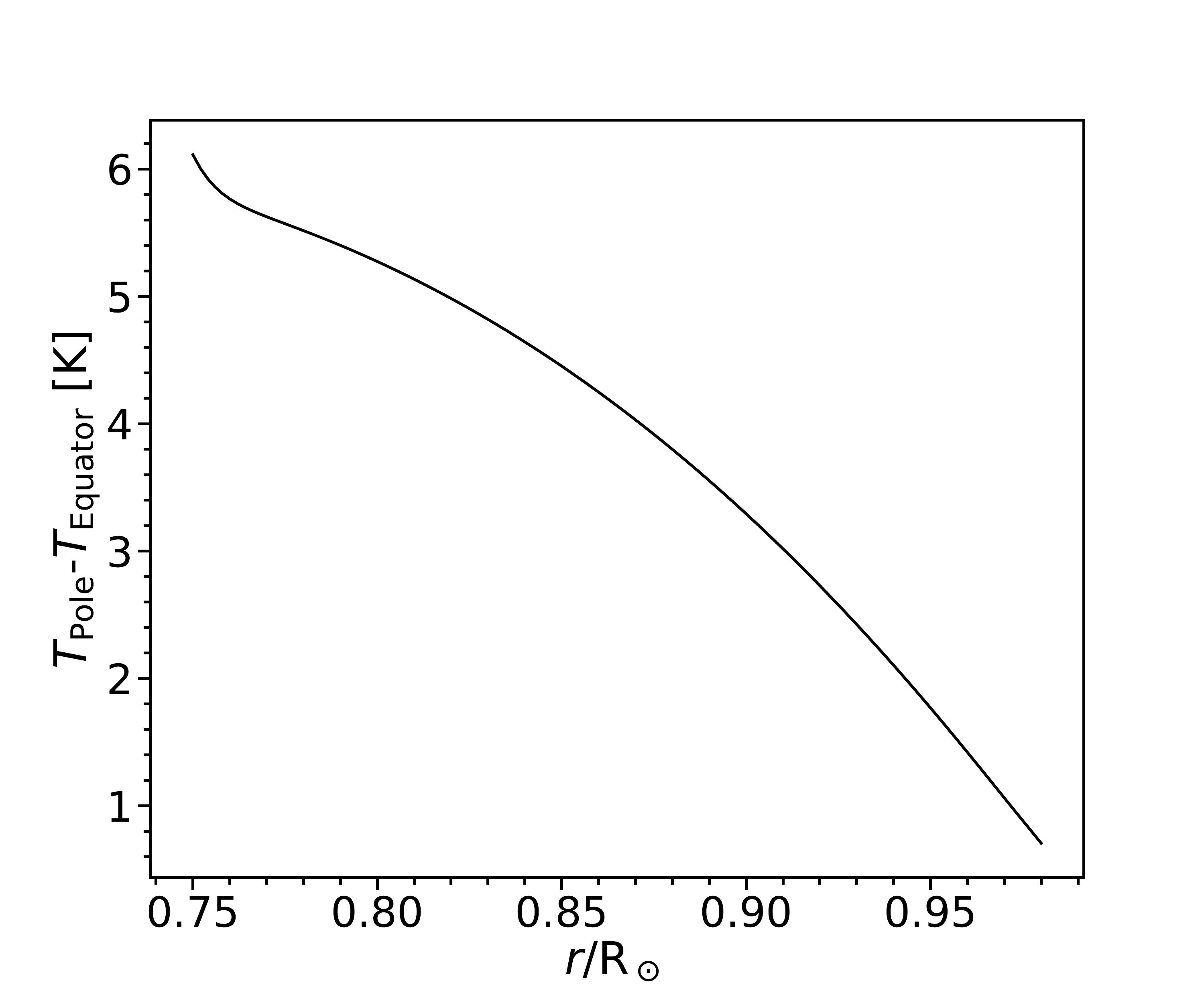}
    \caption{The pole-equator temperature difference as a function of radius, corresponding to Figure~\ref{fig:delta_T_emp}.}
    \label{fig:dt_vs_r_emp}
\end{figure}

\section{Results Based on Analytical Expression}
\label{sec:empirical}
As pointed out in Section~\ref{sec:method}, we need the values of $\Omega (r,\theta)$ below $r<r_c$ to start our calculations. 
In this section, we present the results of our calculations based on the following analytical expression of $\Omega (r,\theta)$ which fits the helioseismology observations closely \citep{Schou1998, Charbonneau1999}:
\begin{equation}
    \Omega(r,\theta)= \Omega_{\rm RZ}+ \frac{1}{2}\left[1-{\rm erf}\left(\frac{r-r_t}{d_t}\right)\right] \left[\Omega_{\rm SCZ}(\theta)-\Omega_{\rm RZ}\right],
    \label{eq6}
\end{equation}
 where $r_t=0.7$\Rsun, $d_t=0.025$\Rsun, $\Omega_{\rm RZ}/2\pi=432.8$ nHz and  $\Omega_{\rm SCZ}(\theta)/2\pi=\Omega_{\rm EQ}+\alpha_2\cos^2(\theta)+\alpha_4\cos^4(\theta)$, with $\Omega_{\rm EQ}/2\pi= 460.7$ nHz, $\alpha_2/2\pi=-62.69$ nHz and $\alpha_4/2\pi=-67.13$ nHz.  Figure~\ref{fig:rot_prof_emp} shows the rotation profile obtained from Equation~\ref{eq6} along with contours of constant $\Omega$ (solid black lines). We note the absence of any signature of NSSL. On comparing with Figure~\ref{fig:omega_helios} giving the rotation profile based on helioseismology data, we see that the analytical expression gives a
 reasonable fit to the data in the deeper layers of the convection zone.
 

\begin{figure*}
	\includegraphics[width=\textwidth]{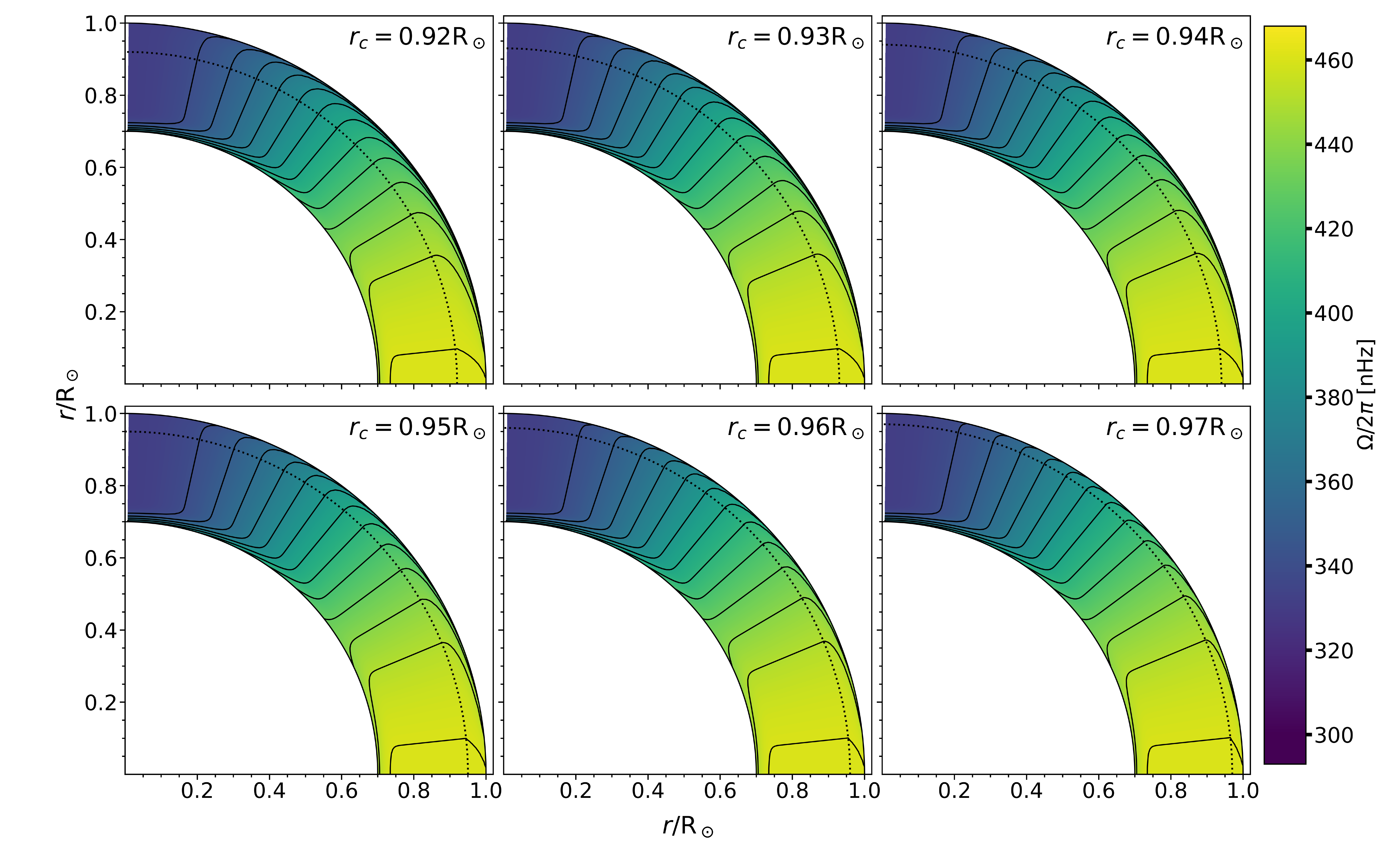}
   \caption{The profiles of $\Omega (r,\theta)$ for different values of $r = r_c$, obtained by using $\Omega (r,\theta)$ given
by Equation~\ref{eq6} as input data for $r < r_c$ .}
    \label{fig:nssl_emp}
\end{figure*}

As explained in Section~\ref{sec:method}, our first step is to obtain $\Delta T (r, \theta)$ for $r<r_c$ by making use of
Equation~\ref{eq3}, in which the left hand side is evaluated by using $\Omega (r, \theta)$ as given by Equation~\ref{eq6}.
To calculate the left side of Equation~\ref{eq3}, we need to evaluate the derivative of $\Omega^2$ along the $z$ direction. For this purpose, we use the transformation equation
\begin{equation}
    \left(\frac{\partial}{\partial z}\right)_s=\left(\frac{\partial r}{\partial z}\right)_s\frac{\partial}{\partial r}-\left(\frac{\partial \theta}{\partial z}\right)_s\frac{\partial}{\partial \theta}=\cos{\theta}\frac{\partial}{\partial r}-\frac{\sin{\theta}}{r}\frac{\partial}{\partial \theta},
    \label{eq7}
\end{equation}
on making use of $s=r\cos{\theta}$ and $z=r\sin{\theta}$ (shown in Figure~\ref{fig:context}). 
We need to choose a particular value of $r_c$. We are going to present discussions for values of $r_c$ in the range 0.92\Rsun\,--\,0.97\Rsun. 
We now use Equation~\ref{eq34} to calculate $\Delta T (r,\theta)$ in the convection zone for all values of $r$ below the maximum value 0.98\Rsun of $r_c$ that we consider. We calculate the numerical derivative of $\Omega^2$ with the help of the transformation Equation~\ref{eq7} by using the first order divided difference scheme.   Then we use the Runge-Kutta 4th order (RK4) method to solve Equation~\ref{eq3}, which is equivalent to carrying on the integration in Equation~\ref{eq34}.

Figure~\ref{fig:delta_T_emp} shows the distribution of $\Delta T (r, \theta)$ in the convection zone below 0.98\Rsun that would follow on assuming the thermal wind balance
and using $\Omega (r, \theta)$ given by the analytical expression Equation~\ref{eq6}. We clearly see a decrease in  $\Delta T(r,\theta)$ with $r$ as we approach the surface. 
Now, one quantity which is of particular interest to us is the pole-equator temperature difference $\Delta T(r, \theta=0) - \Delta T(r, \theta=\pi/2)$ as a function
of $r$. Figure~\ref{fig:dt_vs_r_emp} shows this pole-equator temperature difference as a function of $r$ within the convection zone for $r<0.98$\Rsun.

We shall now present our results for the NSSL by assuming different values of $r_c$ in the range 0.92\Rsun\,--\,0.97\Rsun. For a particular value of
$r_c$, we take $\Delta T (r,\theta)$ to be as given in Figure~\ref{fig:delta_T_emp} for $r <r_c$ and as given by Equation~\ref{eq5} for $r >r_c$.  In this way, we obtain $\Delta T (r,\theta)$
throughout the convection zone for a chosen value of $r_c$.
We point out that the observational value
of the pole-equator temperature difference reported by \citet{rast_2008} is $\approx 2.5$ K. Figure~\ref{fig:dt_vs_r_emp} shows that such a value of the pole-equator temperature difference
occurs at around $r \approx 0.92$\Rsun when we evaluate $\Delta T (r, \theta)$ from the analytical expression (Equation~\ref{eq6}) of $\Omega(r, \theta)$.
This means that we have to take $r_c \approx 0.92$\Rsun to get the pole-equator temperature difference at the surface which matches the 
observations of \citet{rast_2008}.

Now that we have $\Delta T (r,\theta)$ throughout the convection zone including the top layer ($r\ge r_c$) for different values of $r_c$, the last step is to calculate $\Omega (r\ge r_c,\theta)$ in this top layer by assuming that the thermal wind balance holds in this layer. We have already made use of $\Omega (r < r_c,\theta)$ in the deeper layers of the convection zone, as given by Equation~\ref{eq6}, to calculate $\Delta T (r,\theta)$ by making use of Equation~\ref{eq34} (which is effectively
the same as Equation~\ref{eq3}) and  Equation~\ref{eq5}. We now use the thermal balance Equation~\ref{eq3} in the top layer ($r\ge r_c$) in a different manner. From the value of $\Delta T (r\ge r_c,\theta)$ in this top layer, we calculate the right hand side of Equation~\ref{eq3} and then solve Equation~\ref{eq3} to find the distribution of $\Omega (r\ge r_c,\theta)$ in this top layer which would satisfy Equation~\ref{eq3}. We carry on this procedure for the values $r_c = $ 0.92\Rsun, 0.93\Rsun, 0.94\Rsun, 0.95\Rsun, 0.96\Rsun, 0.97\Rsun. We can combine $\Omega (r\ge r_c,\theta)$ obtained in the top layer in this manner with $\Omega (r < r_c,\theta)$ in the deeper layers as given by Equation~\ref{eq6}. This combination for the different values of $r_c$ which we have used are shown in Figure~\ref{fig:nssl_emp}. The dotted circles in the various sub-figures indicate the values of $r_c$ for all these cases.

The contours of constant $\Omega$ bend towards the equator in the top layers of the convection zone for all values of $r_c$ shown in Figure~\ref{fig:nssl_emp}. This indicates the clear presence of the NSSL in all these cases.  We stress again that the initial input data $\Omega (r < r_c,\theta)$ which we had used in the deeper layers of the convection zone in order to start our calculations did not have the NSSL.  In fact, the analytical expression (Equation~\ref{eq6}), which we had used for the values of $\Omega (r < r_c,\theta)$ in the deeper layers, does not give rise to the NSSL at all, as seen in Figure~\ref{fig:rot_prof_emp}. It is thus clear that the NSSL that we see in Figure~\ref{fig:nssl_emp} could not be an artifact of the input data.  The NSSL arises from the fact that the thermal wind term becomes very large in the top layers due to the falling temperature there and the centrifugal term also has to become very large to balance it.  This requirement for satisfying the thermal wind condition (Equation~\ref{eq3}) in the top layer can only be met if there is an NSSL.  We propose this as the explanation for the existence of the NSSL in the top layer of the solar convection zone. The different sub-plots in Figure~\ref{fig:nssl_emp} show that the increase in $r_c$ causes the NSSL to be confined to an increasingly narrower layer near the solar surface.

\section{Results Based on Helioseismology Data}
\label{sec:helio}

After presenting the results based on the analytical expression Equation~\ref{eq6} of $\Omega (r, \theta)$ in the previous section (Section~\ref{sec:empirical}),
we now carry on exactly the same calculations based on the value of $\Omega (r, \theta)$ as given by helioseismology. We use $\Omega (r, \theta)$ averaged
over cycle~23, as supplied to us by H.M.\ Antia. The methodology that was used for
obtaining the $\Omega (r, \theta)$ profile from helioseismology data has been described
by \citet{Antia1998, Antia2008}.
 Our calculations are based on the tabulated value of temporally averaged  $\Omega (r, \theta)$ for all $r$ in the range 0.7\Rsun to \Rsun at steps of 0.005\Rsun and for all $\theta$ in the range of $2^{\circ}$ to $90^{\circ}$ ($88^{\circ}$ to $0^{\circ}$ latitude) at steps of $2^{\circ}$. The profile of $\Omega (r, \theta)$ with the contours of constant $\Omega$ (represented as black solid lines) has been shown in Figure~\ref{fig:omega_helios}.

\begin{figure}
	\includegraphics[width=\columnwidth]{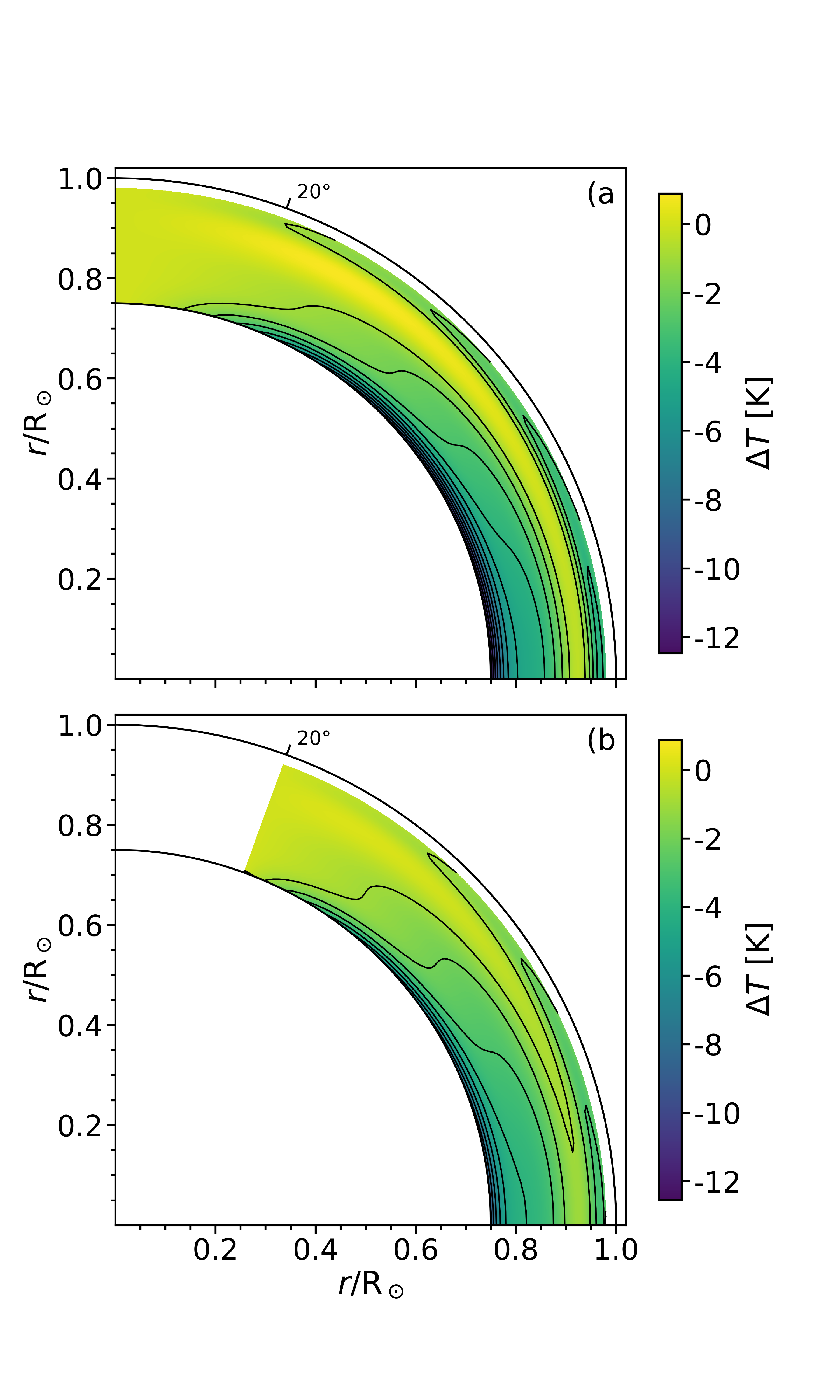}
    \caption{Profile of $\Delta T (r,\theta)$ in the solar convection zone below $r = 0.98$\Rsun, which would follow from the helioseismology data of differential rotation
    on taking $r_c = 0.98$\Rsun. Contours represent the constant values of $\Delta T (r,\theta)$.
     (a) shows the profile of 
    $\Delta T_{0^{\odot}} (r, \theta)$ and (b) the profile
    of $\Delta T_{20^{\odot}} (r, \theta)$ as defined in
    Equation~\ref{eq34}}
    \label{fig:delta_t_helios}
\end{figure}

As in Section~\ref{sec:empirical}, we carry on calculations for different values of $r_c$ in the range 0.93\Rsun\,--\,0.98\Rsun. For a particular value of $r_c$, we substitute the values of $\Omega (r < r_c,\theta)$ in the left hand side of Equation~\ref{eq3} to calculate $\Delta T (r < r_c,\theta)$. The values of $\Delta T (r,\theta)$ for $r > r_c$ are again given by Equation~\ref{eq5}. Figure~\ref{fig:delta_t_helios}a shows the profile of $\Delta T(r<r_c,\theta)$ calculated for the case $r_c = 0.98$\Rsun. One concern we have is that the helioseismic determination of
$\Omega (r, \theta)$ has large uncertainties in the polar regions
at high latitudes and, when we use Equation~\ref{eq34} to calculate $\Delta T (r < r_c,\theta)$, which is $\Delta T_{\theta_0} (r < r_c,\theta)$ with
$\theta_0 = 0$, we have to integrate over this region where the
value of $\Omega (r, \theta)$ is unreliable. One way of avoiding
this difficulty is to consider temperature variations only in 
regions not too close to the poles where we can trust the helioseismic
values of $\Omega(r, \theta)$. We have used Equation~\ref{eq34} to calculate
$\Delta T_{20^{\circ}} (r < r_c,\theta)$ by avoiding the polar
region. Figure~\ref{fig:delta_t_helios}b shows the profile of
$\Delta T_{20^{\circ}} (r < r_c,\theta)$. Comparing the profiles
of $\Delta T (r < r_c,\theta)$ for co-latitudes higher than $20^{\circ}$ (i.e.\ latitudes lower than $70^{\circ}$)
in Figures~\ref{fig:delta_t_helios}a\,--\,b, we find that various
features are in broad agreement, indicating that they are not due
to errors in $\Omega (r, \theta)$ near the polar region. 
Especially, we find an annular strip near $r = 0.925 R_{\odot}$ within which the value of $\Delta T (r < r_c,\theta)$ is close to zero in Figure~\ref{fig:delta_t_helios}a.  This strip becomes
less prominent in Figure~\ref{fig:delta_t_helios}b,
though it does not disappear. The reason behind this strip
is this.  On taking a careful look at Figure~\ref{fig:omega_helios}, we realize
that $\pa \Omega^2/\pa z$ just below the NSSL is close
to zero at latitudes higher than mid-latitudes and is
even positive at very high latitudes (it is usually negative within the convection zone). This explains, on
the basis of Equation~\ref{eq34}, why we have this unusual strip even in
Figure~\ref{fig:delta_t_helios}b after excluding the
polar region. \citet{Schou1998} refer to this region
at high latitudes somewhat below the surface
as ``a submerged polar jet'' and comment in Section~5.5 of
their paper that it ``is seen consistently by several independent methods''.  If this submerged polar jet is
real and not a data artifact, then what causes it is certainly an important
question.  We do not attempt to address this question in
the present paper.

\begin{figure}
	\includegraphics[width=\columnwidth]{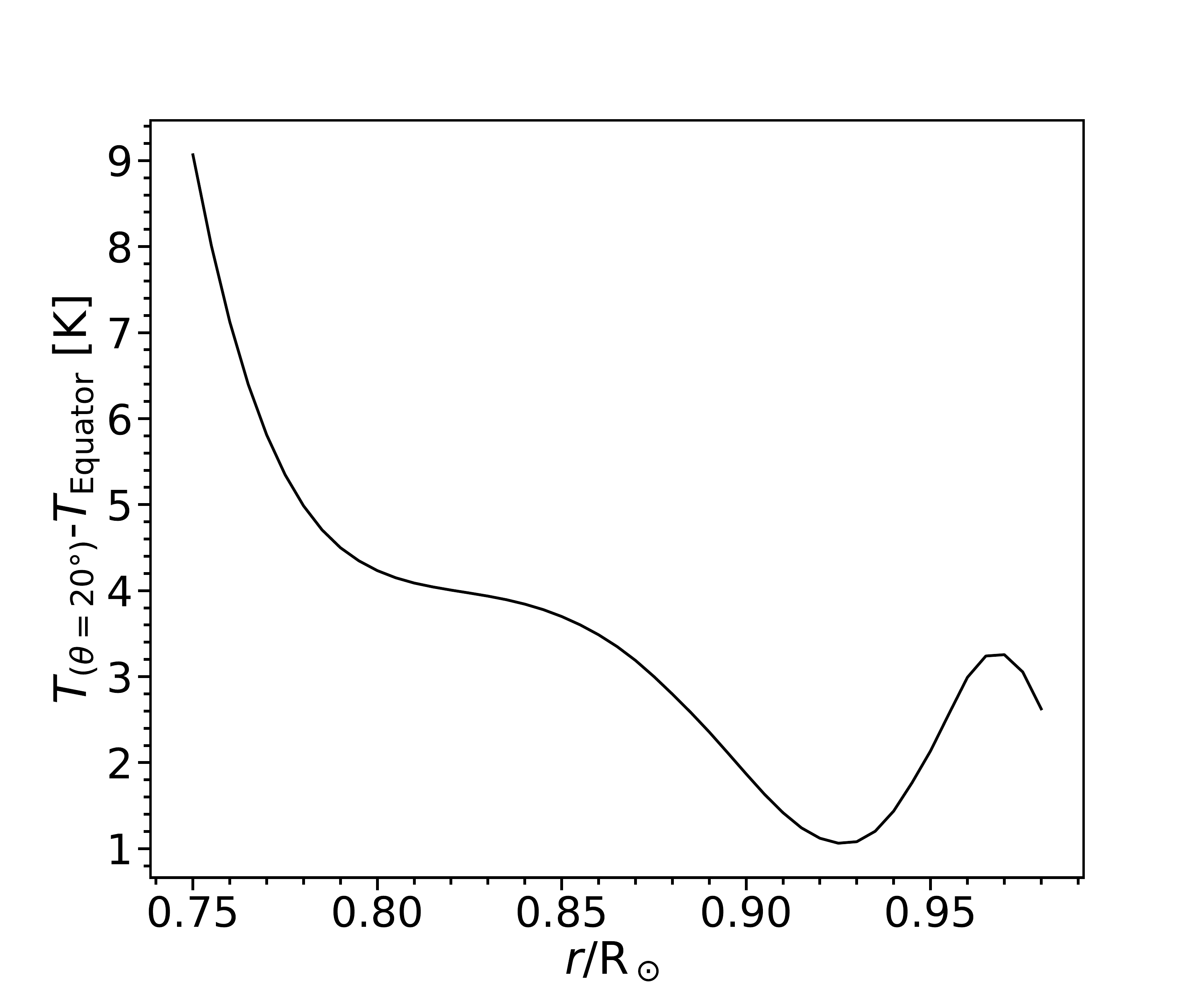}
    \caption{The temperature difference between
    the co-latitude $20^{\circ}$ and the equator
    as a function of radius, corresponding to Figure~\ref{fig:delta_t_helios}.}
    \label{fig:dt_vs_r_helios}
\end{figure}

We now plot the temperature difference $\Delta T (r, \theta=20^{\circ}) - \Delta T (r, \theta=90^{\circ})$ between the 
co-latitude $20^{\circ}$ (i.e.\ latitude $70^{\circ}$) and the
equator as a function of $r$ in Figure~\ref{fig:dt_vs_r_helios}. It should be clear from
Equation~\ref{eq34} that this temperature difference is given by integrating the integrand in the right hand side of Equation~\ref{eq34}
from $\theta=20^{\circ}$ to $\theta=90^{\circ}$. In 
other words, this temperature difference is independent
of the values of $\Omega (r, \theta)$ at very high latitudes (where these values may have large uncertainties) and should be the same for both the cases shown in 
Figure~\ref{fig:delta_t_helios}a and 
Figure~\ref{fig:delta_t_helios}b.  We have seen in the calculations based on the analytical expression of $\Omega (r, \theta)$ in the previous section (Section~\ref{sec:empirical}) that the pole-equator temperature difference decreased monotonically with $r$ (see Figure~\ref{fig:dt_vs_r_emp}). However, Figure~\ref{fig:dt_vs_r_helios} shows a more complicated dependence of a similar temperature difference on $r$.  We indeed find a monotonic decrease of the temperature difference with $r$ for values of $r$ lower than $\approx0.92$\Rsun. But then it starts increasing with $r$ up to $\approx 0.97$\Rsun, beyond which it decreases again. This complicated variation of the temperature difference is  connected with the submerged polar jet which continues even a little bit beyond co-latitude $20^{\circ}.$

\begin{figure*}
	\includegraphics[width=\textwidth]{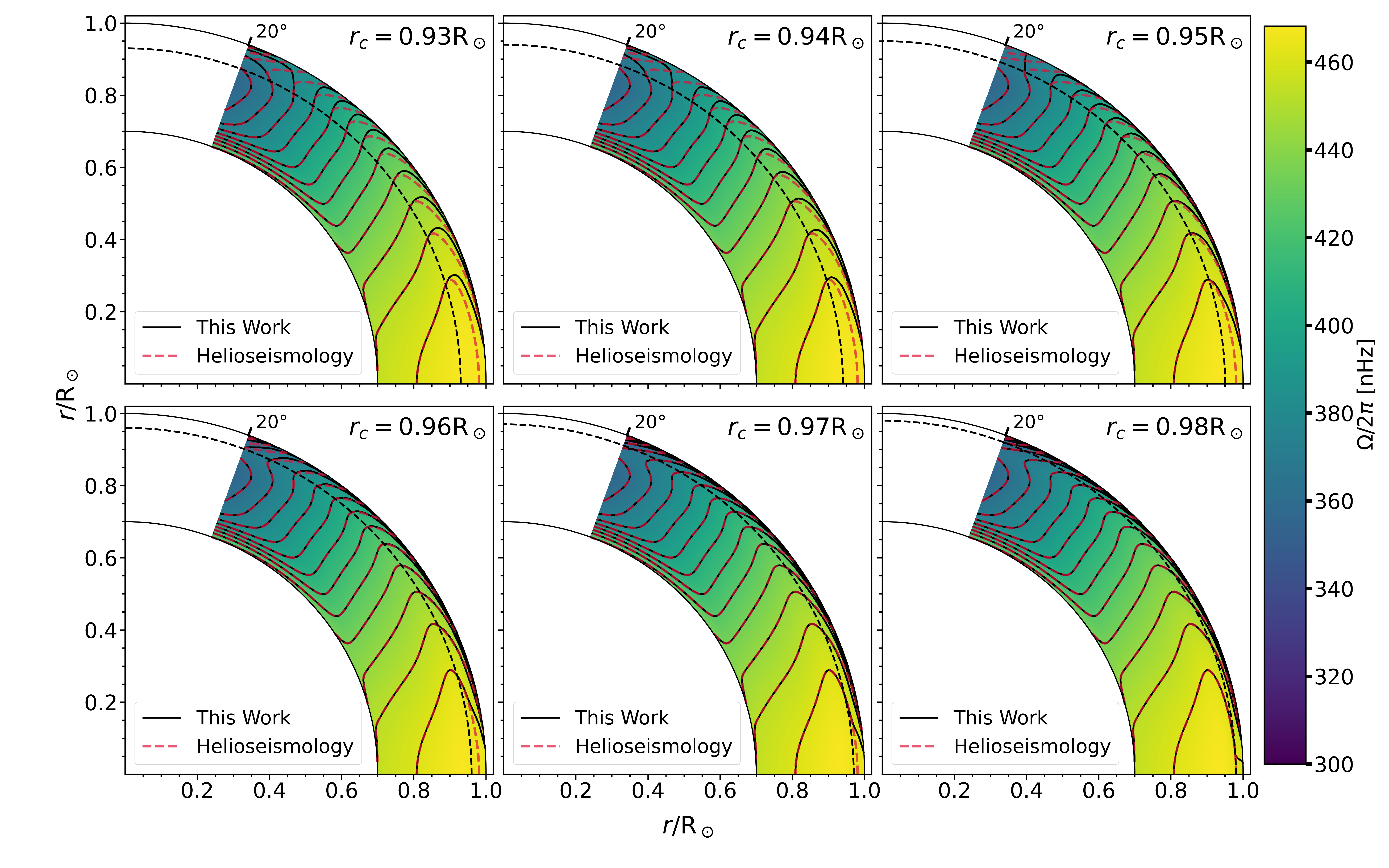}
    \caption{The profiles of $\Omega (r,\theta)$ for different values of $r = r_c$, obtained by using $\Omega (r,\theta)$ given by helioseismology as input data for $r < r_c$ .}
    \label{fig:nssl_helios}
\end{figure*}

The final step in our analysis is exactly the same as in Section~\ref{sec:empirical}.  Once we have $\Delta T (r, \theta)$ throughout the convection zone corresponding to different values of $r_c$ (with its value for $r \ge r_c$ being given by Equation~\ref{eq5}), we evaluate the right hand side of Equation~\ref{eq3} for $r\ge r_c$ and then solve Equation~\ref{eq3} to obtain $\Omega (r\ge r_c, \theta)$ in the top layers of the convection zone. Since we have to differentiate $\Delta T (r > r_c,\theta)$  with respect to $\theta$, it does not matter whether we use
$\Delta T_{\theta_0} (r > r_c,\theta)$ with $\theta_0 = 0$ or with
$\theta_0 = 20^{\circ}$. The temperature profiles in both 
Figure~\ref{fig:delta_t_helios}a and Figure~\ref{fig:delta_t_helios}b give the same distribution
of $\Omega (r\ge r_c, \theta)$ in the top layers of the convection
zone for $\theta$ higher than $20^{\circ}$. Thus, the profile
of $\Omega (r\ge r_c, \theta)$ in the top layers of the convection
zone that we have calculated starting initially from $\Omega (r\le r_c, \theta)$
in the deeper layers of the convection is independent of the errors
in $\Omega (r\le r_c, \theta)$ in the polar region.
In Figure~\ref{fig:nssl_helios}, we have shown the distribution of $\Omega (r\ge r_c, \theta)$ obtained in this way for different values of $r_c$ (represented by black dashed circles), along with $\Omega (r<r_c,\theta)$ as given by helioseismology data (same as in Figure~\ref{fig:omega_helios}). In all these cases, we clearly see the NSSL. While the input data $\Omega (r < r_c, \theta)$ used in our calculations show some indications of the NSSL for the cases $r_c = $0.97\Rsun, 0.98\Rsun, there was no sign of the NSSL 
in the input data for the cases $r_c = $0.93\Rsun, 0.94\Rsun. The fact that we get a layer just below the solar surface resembling the NSSL in all these cases strongly suggests that the NSSL arises from the requirement of the thermal wind balance with the thermal wind term becoming very large in the top layer of the convection zone. To facilitate comparison of our theoretical results with the observations, we have over-plotted in Figure~\ref{fig:nssl_helios} the contours of constant $\Omega$ obtained from helioseismology observation (shown by dashed red lines).

\begin{figure}
	\includegraphics[width=\columnwidth]{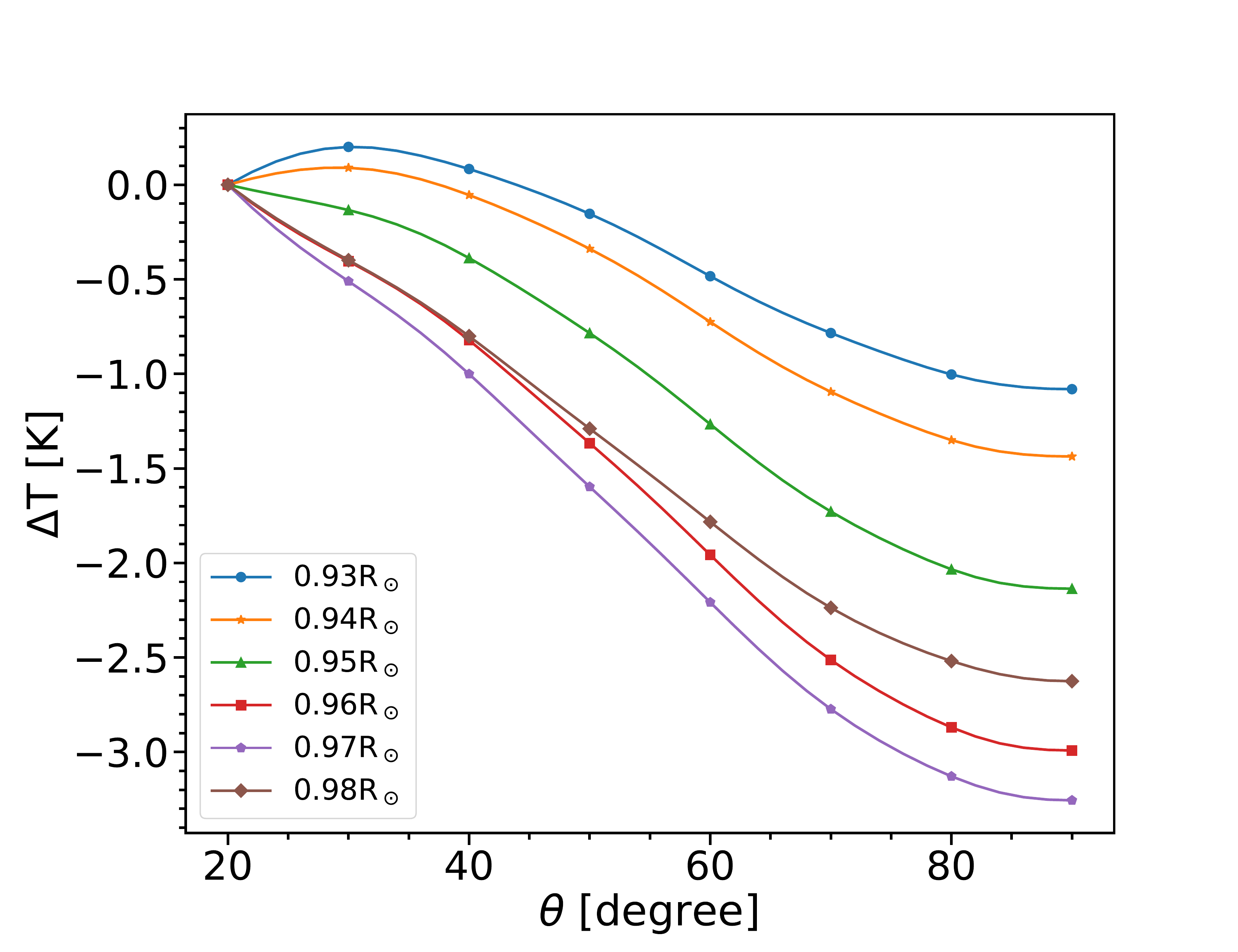}
    \caption{Variation of $\Delta T_{20^{\circ}} (R_\odot, \theta)$ on the solar surface with co-latitude $\theta$ for different values of $r_c$ (markers are plotted at every 5 data points).}
    \label{fig:dt_theta}
\end{figure}

We at last come to the question whether the theoretical results obtained on the basis of our assumption that the thermal balance equation holds in the top layers of the convection zone agree with observational data.  Comparing the solid lines indicating the theoretical results with the dashed red lines indicating helioseismology data above $r = r_c$ in Figure~\ref{fig:nssl_helios}, it is evident that the agreement is very good for all values of $r$ larger than 0.96\Rsun. In the cases $r = $0.97\Rsun, 0.98\Rsun, the lower part of the NSSL was present in the input data and one may argue that it is not so surprising that our theoretical calculations correctly gave the structure of the NSSL in the upper layers.  However, this is clearly not case for $r = 0.96$\Rsun. One other aspect of the observational data we need to match is the pole-equator temperature difference. Looking at Figure~4 of \citet{rast_2008}, we find that they present measurements
up to latitudes of about $70^{\circ}$.  What they loosely
refer to as the pole-equator temperature difference (PETD)
is actually the temperature difference between $70^{\circ}$ and the equator.  If we also use the same convention,
then what is plotted in Figure~\ref{fig:dt_vs_r_helios}
can be called PETD and compared directly with the results
of \citet{rast_2008}. Taking the value 2.5~K reported by \citet{rast_2008} to be the correct value, we note in Figure~\ref{fig:dt_vs_r_helios} that the PETD has the value 2.5 K at $r \approx 0.96$\Rsun.  If this is taken to be the value of $r_c$, then PETD at the solar surface should also be 2.5 K. It is thus clear that an accurate determination of the PETD at the solar surface is extremely important and can put constraints on the appropriate value of $r_c$ to be used in theoretical calculations.  If this temperature difference is indeed 2.5 K, then we conclude that the theoretical calculations carried out with $r_c=0.96$\Rsun in our model are in good agreement with observational data.  This case gives a good structure of the NSSL as we see in Figure~\ref{fig:nssl_helios} and the PETD at the solar surface also has the desired value 2.5 K.

\begin{figure}
	\includegraphics[width=\columnwidth]{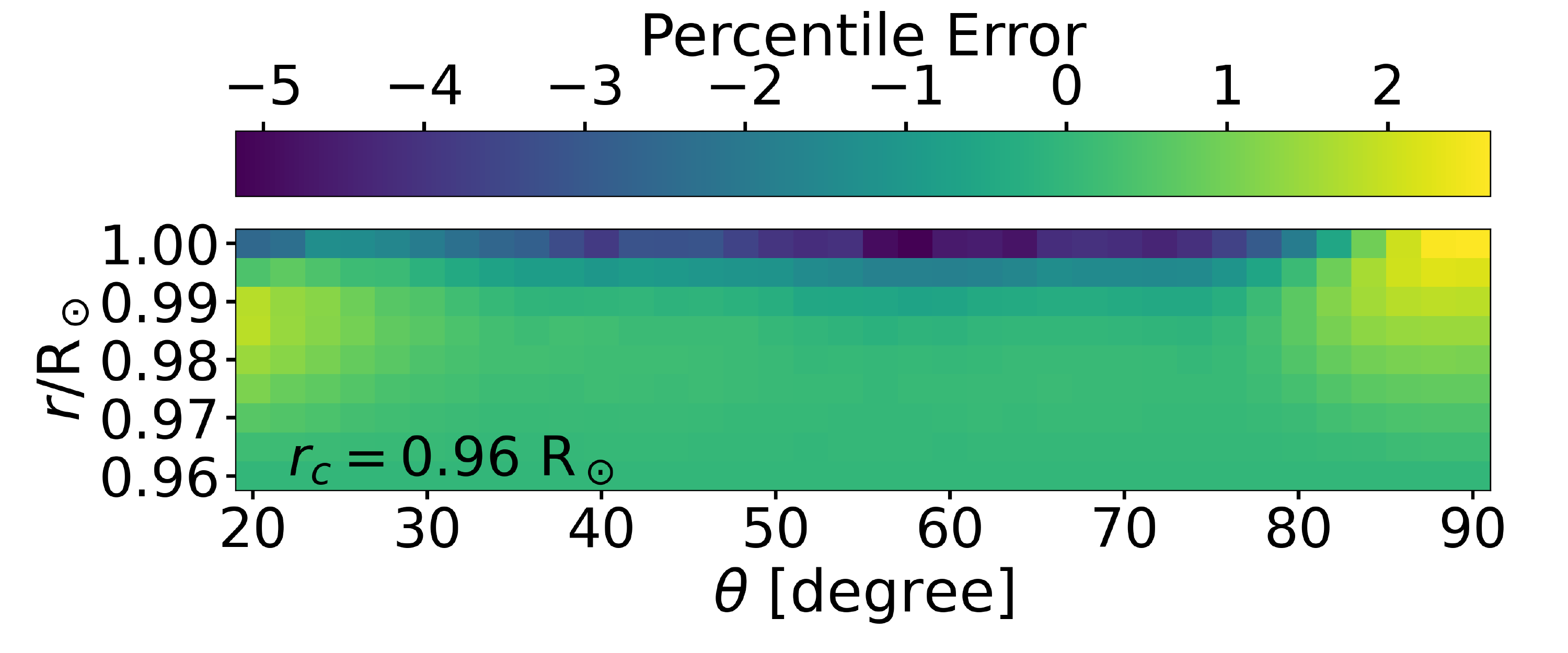}
    \caption{Distribution of percentile error $f$ for the case of $r_c=0.96$ \Rsun.}
    \label{fig:error}
\end{figure}

To check quantitatively how well our theoretically
determined $\Omega$ in the NSSL
compares with observational  $\Omega_{\rm heliseismology}$, we consider the percentile error
\begin{equation}
    f = 100 \frac{\Omega - \Omega_{\rm helioseismology}}
    {\Omega_{\rm helioseismology}}
    \label{eq11}
\end{equation}
at different points. Since the observed variation of $\Omega$ within the NSSL is at the level of 5\%, we
must have $f$ considerably less than that for the fit
to be considered sufficiently good. Figure~\ref{fig:error} shows the
distribution of $f$ for the case $r_c = 0.96 $\Rsun  in
the top layer of the convection zone above $r = 0.96$ \Rsun. We indeed find that $f$ is much less than
5\% within the NSSL, except in a very thin layer
above $0.995$\Rsun close to the solar surface. This
perhaps suggests that the thermal wind balance breaks
down in this very thin layer near the surface. The smallness
of $f$ within the NSSL below this very thin layer presumably
indicates that the thermal wind balance holds there to
a very good approximation. The root mean square (RMS) value of $f$ for the 
case 
presented in Figure~\ref{fig:error} is found to be 1.36\%. 
The RMS values of $f$ for
cases  $r_c = 0.95 $\Rsun and $r_c = 0.97$\Rsun turn
out to be 1.50\% and 2.41\% respectively.

We are not aware of any independent measurements of the pole-equator temperature  difference after the work of \citet{rast_2008} done more than a decade ago. If the variation of the temperature with latitude on the solar surface is be measured more accurately by modern techniques in the future, then it will be useful to compare such observations with the theoretical results of our model. Figure~\ref{fig:dt_theta} shows how 
$\Delta T_{20^{\circ}} (R_\odot, \theta)$ varies with the co-latitude $\theta$ for different values of $r_c$.  Note that $\Delta T_{20^{\circ}} (r, \theta)$ is defined in Equation~\ref{eq34} in such a manner that its value is always zero at $\theta = 20^{\circ}$. Also, note that the curves for cases $r_c = 0.96$ \Rsun, 0.97 \Rsun, 0.98 \Rsun  do not appear in the same simple sequence as the curves for cases $r_c = 0.93$ \Rsun, 0.94 \Rsun, 0.95 \Rsun . Given the complicated plot shown in Figure~\ref{fig:dt_vs_r_helios}, this behaviour is not surprising.  It will be instructive to compare Figure~\ref{fig:dt_theta} with observational data when such data become available.  We remind the readers that all the results in this Section were obtained on the assumption of a sudden jump in the nature of convective heat transport at $r = r_c$. If the transition is more gradual, that may change the results slightly.  Detailed comparison between theoretical results and observational data in future may throw more light on this.

Assuming that the intensity of radiation $I$ emitted from a region of the
surface goes as $T^4$ according to the Stefan-Boltzmann law, the variation
of intensity with latitude caused by the variation of temperature with latitude would be given by
\begin{equation}
    \frac{\Delta I}{I} \approx 4 \frac{\Delta T}{T}.
\end{equation}
The pole-equator intensity difference corresponding to a temperature difference of 2.5 K would be
\begin{equation}
    \frac{\Delta I}{I} \approx 0.0016.
\end{equation}
In other words, the pole would be only about 0.16\% brighter than the
equator. The latitudinal variation of intensity measured by \citet{rast_2008}
is given in Figure~4 of their paper.  They point out that it is non-trivial
to measure this small latitudinal variation of intensity.  Apart from 
instrumental errors, the presence of polar faculae makes these measurements
difficult. However, \citet{rast_2008} claimed that their measurement of 
enhanced intensity near the polar region is a real physical effect.

\section{Conclusion}
\label{sec:conclusion}
There are differences of opinion why the Sun has a Near-Surface Shear Layer (NSSL).  A novel explanation of the NSSL was recently proposed by \citet{Choudhuri2021a} on the basis of order-of-magnitude estimates.  We now substantiate the proposal of \citet{Choudhuri2021a} through detailed  calculations. Although it is generally agreed that the thermal wind balance equation holds within the body of the solar convection zone, whether this equation even holds in the top layers had been debated. It has been suggested that this equation breaks down in a boundary layer at the top and this may somehow give rise to the NSSL. \citet{Choudhuri2021a} argued on the other hand that the thermal wind term becomes very large in the top layers of the convection zone and the thermal wind balance equation has to hold. This means that the centrifugal term also has to be very large in the top layers to achieve the thermal wind balance, necessitating the existence of the NSSL. When we compare the final results of our
theoretical model with observational data from helioseismology, we conclude that the thermal wind
balance condition may break down only in a very thin
layer near the solar surface (having thickness of order $\approx 0.005$\Rsun or
$\approx 3000$ km).

The argument proposed in this paper hinges crucially on the fact that the convective cells are affected by the solar rotation within the main body of the convection zone, except in the top layers within which the convective turnover time is less than the rotation period of the Sun.  Although the transition from the layers where rotational effects are large to the layers where they are small must be a gradual transition, we simplify the calculations by assuming that this transition takes place at $r = r_c$. Since the effect of the solar rotation would make the convective heat transport in the deeper layers dependent on latitude, we expect that the temperature at a point within the convection zone will depart by an amount $\Delta T (r, \theta)$ from what we get from the standard models of the convection zone in which the effect of rotation is not taken into account.  In principle, it should be possible to calculate $\Delta T (r, \theta)$ from the theory of convective heat transport.  However, this is a formidably difficult problem in practice and we calculate $\Delta T (r, \theta)$ in the deeper layers of the convection zone from the differential rotation measured by helioseismology by assuming that the thermal wind balance condition prevails in the deep layers of the convection zone. Since the effect of rotation on the convection is negligible above $r = r_c$, we expect the radial temperature gradient $dT/dr$ to be independent of latitude in this layer, which implies that $\Delta T (r, \theta)$ does not vary with $r$ in this layer.  It is this fact, coupled with the fact that the temperature drops sharply in this top layer, which makes the thermal wind term very large in this top layer.  From the requirement that the centrifugal term also has to become large to balance the large thermal wind term in this layer, we can calculate the distribution of $\Omega (r, \theta)$ in this near-surface layer.  We have found that our calculations give a layer resembling the NSSL.

The non-variation of $\Delta T (r, \theta)$ with $r$ in the upper layers of the convection zone ensures that the pole-equator temperature difference does not vary in this layer.  The means that the value of the pole-equator temperature difference at $r_c$ gets mapped to the solar surface.  A careful measurement of the pole-equator temperature difference at the solar surface would enable us to assess the value of $r_c$ above which the convective motions are not affected much by rotation.  The value 2.5 K of the pole-equator temperature difference reported by \citet{rast_2008} led us to conclude that $r_c \approx 0.96$\Rsun.  For this value of $r_c$, the various aspects of observational data, including the structure of the NSSL, are explained very well by our theoretical model. Fairly sophisticated simulations of solar convection are now being carried on by many groups. We hope that such
simulations may also eventually be able to give an indication of the value of $r_c$ above which the effect of rotation is negligible.  
Since the measurement of the pole-equator temperature difference allows us to assess $r_c$, such a measurement can put important constraints on the simulations of solar convection. It appears that there have not been any independent measurements of the pole-equator temperature  difference after the work of \citet{rast_2008} done more than a decade ago.  We hope that other groups will undertake this measurement in the near future, since the value of this temperature difference has connections with such important issues as the nature of the solar convection and the structure of the NSSL.

The large-scale flows in the solar convection zone like the differential rotation and the meridional circulation play important roles in the flux transport dynamo model for explaining the solar cycle, which started being developed from the 1990s \citep{Wang1991,Choudhuri1995, Durney1995} and has been reviewed by several authors in the last few years \citep{Charbonneau2010, Choudhuri2011, karak_2014}.  One crucial question is whether the NSSL is important in the solar dynamo process.  One key idea in the solar dynamo models is that the toroidal magnetic field is generated by the strong differential rotation at the bottom of the solar convection zone, where the field can be stored in the stable sub-adiabatic layers below the bottom of the convection zone and can undergo amplification there. Srtands of the toroidal magnetic field eventually break out of the stable layers to rise through the convection zone due to magnetic buoyancy. Since the near-surface layer is a region of strong super-adiabatic temperature gradient which enhances magnetic buoyancy \citep{Moreno1983, Choudhuri1987}, magnetic fields are expected to rise through this layer quickly without allowing much time for shear amplification. Unless there is some mechanism to keep magnetic fields stored in the NSSL for some time, most likely the NSSL is not important for the dynamo process,
although there is not complete unanimity on this  \citep{Brandenburg2005}.
Models of the flux transport dynamo without including the NSSL give reasonable fits with observations \citep{Chatterjee2004}. 
Dynamo-generated magnetic fields, however, can react back on the large-scale flows producing temporal variations with the solar cycle \citep{chakraborty_2009, hazra_2017}. For example, the meridional circulation varies periodically with the solar cycle and modelling it requires going beyond the thermal wind balance Equation~\ref{eq3} to include a time derivative term \citep{hazra_2017, Choudhuri2021b}. The thermal wind balance equation follows from the full equation for the meridional circulation under steady state conditions if the dissipation term can be ignored.  Presumably, the thermal wind balance Equation~\ref{eq3} holds for the time-averaged part of large-scale flows, which has been our focus in this paper. However, there is evidence of random temporal fluctuations in the meridional circulation \citep{karak2010, karak2011, Choudhuri2012, Choudhuri2014, Hazra2019}, possibly indicating slight violations of the thermal wind balance equation. Since the terms involved in the thermal wind balance are much larger than the other terms in the equation of the meridional circulation (see, for example, the discussion in the Appendix of \citet{Choudhuri2021b}), even a slight imbalance between these large terms is sufficient to cause fluctuation in the meridional circulations and we believe that the violations of the thermal wind balance remain very small.

Lastly, we suggest that other solar-like stars with rotation periods similar to the Sun are likely to have similar shear layers near their surfaces, since the solar NSSL arises out of very general considerations which should hold for such stars.  The study of starspots and stellar cycles in the last few years have suggested that the solar-like stars also must have large-scale flows like the differential rotation and the meridional circulation giving rise to dynamo cycles, as in the case of the Sun \citep{karak2014b, Choudhuri2017, Hazra2019b}.  Although asteroseismology has started giving some initial results of differential rotation in solar-like stars \citep{Benomar2018}, we are still very far for determining observationally whether other stars also have NSSL.

\section*{Acknowledgements}
We thank Dipankar Banerjee for useful discussions throughout this work and H.M.\ Antia for providing the helioseismology data that we needed for our analysis. Extremely valuable suggestions from an anonymous referee helped in improving the manuscript significantly.

\section*{DATA AVAILABILITY}
The helioseismology data utilized in this article were supplied to us by Prof.\ H.M.\ Antia. The data underlying this work will be shared on reasonable request to the corresponding author.




\bibliographystyle{mnras}

\begin{thebibliography}{}
    \makeatletter
    \relax
    \def\mn@urlcharsother{\let\do\@makeother \do\$\do\&\do\#\do\^\do\_\do\%\do\~}
    \def\mn@doi{\begingroup\mn@urlcharsother \@ifnextchar [ {\mn@doi@}
      {\mn@doi@[]}}
    \def\mn@doi@[#1]#2{\def\@tempa{#1}\ifx\@tempa\@empty \href
      {http://dx.doi.org/#2} {doi:#2}\else \href {http://dx.doi.org/#2} {#1}\fi
      \endgroup}
    \def\mn@eprint#1#2{\mn@eprint@#1:#2::\@nil}
    \def\mn@eprint@arXiv#1{\href {http://arxiv.org/abs/#1} {{\tt arXiv:#1}}}
    \def\mn@eprint@dblp#1{\href {http://dblp.uni-trier.de/rec/bibtex/#1.xml}
      {dblp:#1}}
    \def\mn@eprint@#1:#2:#3:#4\@nil{\def\@tempa {#1}\def\@tempb {#2}\def\@tempc
      {#3}\ifx \@tempc \@empty \let \@tempc \@tempb \let \@tempb \@tempa \fi \ifx
      \@tempb \@empty \def\@tempb {arXiv}\fi \@ifundefined
      {mn@eprint@\@tempb}{\@tempb:\@tempc}{\expandafter \expandafter \csname
      mn@eprint@\@tempb\endcsname \expandafter{\@tempc}}}
    
    \bibitem[\protect\citeauthoryear{{Antia}, {Basu}  \& {Chitre}}{{Antia}
      et~al.}{1998}]{Antia1998}
    {Antia} H.~M.,  {Basu} S.,   {Chitre} S.~M.,  1998, \mn@doi [\mnras]
      {10.1046/j.1365-8711.1998.01635.x}, \href
      {https://ui.adsabs.harvard.edu/abs/1998MNRAS.298..543A} {298, 543}
    
    \bibitem[\protect\citeauthoryear{{Antia}, {Basu}  \& {Chitre}}{{Antia}
      et~al.}{2008}]{Antia2008}
    {Antia} H.~M.,  {Basu} S.,   {Chitre} S.~M.,  2008, \mn@doi [\apj]
      {10.1086/588523}, \href
      {https://ui.adsabs.harvard.edu/abs/2008ApJ...681..680A} {681, 680}
    
    \bibitem[\protect\citeauthoryear{{Bahcall} \& {Pinsonneault}}{{Bahcall} \&
      {Pinsonneault}}{2004}]{Bahcall2004}
    {Bahcall} J.~N.,  {Pinsonneault} M.~H.,  2004, \mn@doi [\prl]
      {10.1103/PhysRevLett.92.121301}, \href
      {https://ui.adsabs.harvard.edu/abs/2004PhRvL..92l1301B} {92, 121301}
    
    \bibitem[\protect\citeauthoryear{{Bahcall} \& {Ulrich}}{{Bahcall} \&
      {Ulrich}}{1988}]{bahcall_1988}
    {Bahcall} J.~N.,  {Ulrich} R.~K.,  1988, \mn@doi [Reviews of Modern Physics]
      {10.1103/RevModPhys.60.297}, \href
      {https://ui.adsabs.harvard.edu/abs/1988RvMP...60..297B} {60, 297}
    
    \bibitem[\protect\citeauthoryear{{Basu}}{{Basu}}{2016}]{Basu2016}
    {Basu} S.,  2016, \mn@doi [Living Reviews in Solar Physics]
      {10.1007/s41116-016-0003-4}, \href
      {https://ui.adsabs.harvard.edu/abs/2016LRSP...13....2B} {13, 2}
    
    \bibitem[\protect\citeauthoryear{{Belvedere} \& {Paterno}}{{Belvedere} \&
      {Paterno}}{1976}]{belvedere_1976}
    {Belvedere} G.,  {Paterno} L.,  1976, \mn@doi [\solphys] {10.1007/BF00154761},
      \href {https://ui.adsabs.harvard.edu/abs/1976SoPh...47..525B} {47, 525}
    
    \bibitem[\protect\citeauthoryear{{Benomar} et~al.,}{{Benomar}
      et~al.}{2018}]{Benomar2018}
    {Benomar} O.,  et~al., 2018, \mn@doi [Science] {10.1126/science.aao6571}, \href
      {https://ui.adsabs.harvard.edu/abs/2018Sci...361.1231B} {361, 1231}
    
    \bibitem[\protect\citeauthoryear{{Brandenburg}}{{Brandenburg}}{2005}]{Brandenburg2005}
    {Brandenburg} A.,  2005, arXiv e-prints, \href
      {https://ui.adsabs.harvard.edu/abs/2005astro.ph.12638B} {pp
      astro--ph/0512638}
    
    \bibitem[\protect\citeauthoryear{{Brown}, {Browning}, {Brun}, {Miesch}  \&
      {Toomre}}{{Brown} et~al.}{2010}]{Brown2010}
    {Brown} B.~P.,  {Browning} M.~K.,  {Brun} A.~S.,  {Miesch} M.~S.,   {Toomre}
      J.,  2010, \mn@doi [\apj] {10.1088/0004-637X/711/1/424}, \href
      {https://ui.adsabs.harvard.edu/abs/2010ApJ...711..424B} {711, 424}
    
    \bibitem[\protect\citeauthoryear{{Brun}, {Antia}  \& {Chitre}}{{Brun}
      et~al.}{2010}]{Brun2010}
    {Brun} A.~S.,  {Antia} H.~M.,   {Chitre} S.~M.,  2010, \mn@doi [\aap]
      {10.1051/0004-6361/200913166}, \href
      {https://ui.adsabs.harvard.edu/abs/2010A&A...510A..33B} {510, A33}
    
    \bibitem[\protect\citeauthoryear{{Chakraborty}, {Choudhuri}  \&
      {Chatterjee}}{{Chakraborty} et~al.}{2009}]{chakraborty_2009}
    {Chakraborty} S.,  {Choudhuri} A.~R.,   {Chatterjee} P.,  2009, \mn@doi [\prl]
      {10.1103/PhysRevLett.102.041102}, \href
      {https://ui.adsabs.harvard.edu/abs/2009PhRvL.102d1102C} {102, 041102}
    
    \bibitem[\protect\citeauthoryear{{Charbonneau}}{{Charbonneau}}{2010}]{Charbonneau2010}
    {Charbonneau} P.,  2010, \mn@doi [Living Reviews in Solar Physics]
      {10.12942/lrsp-2010-3}, \href
      {https://ui.adsabs.harvard.edu/abs/2010LRSP....7....3C} {7, 3}
    
    \bibitem[\protect\citeauthoryear{{Charbonneau}, {Christensen-Dalsgaard},
      {Henning}, {Larsen}, {Schou}, {Thompson}  \& {Tomczyk}}{{Charbonneau}
      et~al.}{1999}]{Charbonneau1999}
    {Charbonneau} P.,  {Christensen-Dalsgaard} J.,  {Henning} R.,  {Larsen} R.~M.,
      {Schou} J.,  {Thompson} M.~J.,   {Tomczyk} S.,  1999, \mn@doi [\apj]
      {10.1086/308050}, \href
      {https://ui.adsabs.harvard.edu/abs/1999ApJ...527..445C} {527, 445}
    
    \bibitem[\protect\citeauthoryear{{Chatterjee}, {Nandy}  \&
      {Choudhuri}}{{Chatterjee} et~al.}{2004}]{Chatterjee2004}
    {Chatterjee} P.,  {Nandy} D.,   {Choudhuri} A.~R.,  2004, \mn@doi [\aap]
      {10.1051/0004-6361:20041199}, \href
      {https://ui.adsabs.harvard.edu/abs/2004A&A...427.1019C} {427, 1019}
    
    \bibitem[\protect\citeauthoryear{Choudhuri}{Choudhuri}{1998}]{Choudhuri1998book}
    Choudhuri A.~R.,  1998, The Physics of Fluids and Plasmas: An Introduction for
      Astrophysicists.
    Cambridge University Press, \mn@doi{10.1017/CBO9781139171069}
    
    \bibitem[\protect\citeauthoryear{{Choudhuri}}{{Choudhuri}}{2011}]{Choudhuri2011}
    {Choudhuri} A.~R.,  2011, \mn@doi [Pramana] {10.1007/s12043-011-0113-4}, \href
      {https://ui.adsabs.harvard.edu/abs/2011Prama..77...77C} {77, 77}
    
    \bibitem[\protect\citeauthoryear{{Choudhuri}}{{Choudhuri}}{2014}]{Choudhuri2014}
    {Choudhuri} A.~R.,  2014, \mn@doi [Indian Journal of Physics]
      {10.1007/s12648-014-0481-y}, \href
      {https://ui.adsabs.harvard.edu/abs/2014InJPh..88..877C} {88, 877}
    
    \bibitem[\protect\citeauthoryear{{Choudhuri}}{{Choudhuri}}{2017}]{Choudhuri2017}
    {Choudhuri} A.~R.,  2017, \mn@doi [Science China Physics, Mechanics, and
      Astronomy] {10.1007/s11433-016-0413-7}, \href
      {https://ui.adsabs.harvard.edu/abs/2017SCPMA..60a9601C} {60, 19601}
    
    \bibitem[\protect\citeauthoryear{{Choudhuri}}{{Choudhuri}}{2021a}]{Choudhuri2021b}
    {Choudhuri} A.~R.,  2021a, \mn@doi [Science China Physics, Mechanics, and
      Astronomy] {10.1007/s11433-020-1628-1}, \href
      {https://ui.adsabs.harvard.edu/abs/2021SCPMA..6439601C} {64, 239601}
    
    \bibitem[\protect\citeauthoryear{{Choudhuri}}{{Choudhuri}}{2021b}]{Choudhuri2021a}
    {Choudhuri} A.~R.,  2021b, \mn@doi [\solphys] {10.1007/s11207-021-01784-7},
      \href {https://ui.adsabs.harvard.edu/abs/2021SoPh..296...37C} {296, 37}
    
    \bibitem[\protect\citeauthoryear{{Choudhuri} \& {Gilman}}{{Choudhuri} \&
      {Gilman}}{1987}]{Choudhuri1987}
    {Choudhuri} A.~R.,  {Gilman} P.~A.,  1987, \mn@doi [\apj] {10.1086/165243},
      \href {https://ui.adsabs.harvard.edu/abs/1987ApJ...316..788C} {316, 788}
    
    \bibitem[\protect\citeauthoryear{{Choudhuri} \& {Karak}}{{Choudhuri} \&
      {Karak}}{2012}]{Choudhuri2012}
    {Choudhuri} A.~R.,  {Karak} B.~B.,  2012, \mn@doi [\prl]
      {10.1103/PhysRevLett.109.171103}, \href
      {https://ui.adsabs.harvard.edu/abs/2012PhRvL.109q1103C} {109, 171103}
    
    \bibitem[\protect\citeauthoryear{{Choudhuri}, {Schussler}  \&
      {Dikpati}}{{Choudhuri} et~al.}{1995}]{Choudhuri1995}
    {Choudhuri} A.~R.,  {Schussler} M.,   {Dikpati} M.,  1995, \aap, \href
      {https://ui.adsabs.harvard.edu/abs/1995A&A...303L..29C} {303, L29}
    
    \bibitem[\protect\citeauthoryear{{Christensen-Dalsgaard}
      et~al.,}{{Christensen-Dalsgaard} et~al.}{1996}]{Dalsgaard1996}
    {Christensen-Dalsgaard} J.,  et~al., 1996, \mn@doi [Science]
      {10.1126/science.272.5266.1286}, \href
      {https://ui.adsabs.harvard.edu/abs/1996Sci...272.1286C} {272, 1286}
    
    \bibitem[\protect\citeauthoryear{{Durney}}{{Durney}}{1995}]{Durney1995}
    {Durney} B.~R.,  1995, \mn@doi [\solphys] {10.1007/BF00732805}, \href
      {https://ui.adsabs.harvard.edu/abs/1995SoPh..160..213D} {160, 213}
    
    \bibitem[\protect\citeauthoryear{{Durney} \& {Roxburgh}}{{Durney} \&
      {Roxburgh}}{1971}]{durney_1971}
    {Durney} B.~R.,  {Roxburgh} I.~W.,  1971, \mn@doi [\solphys]
      {10.1007/BF00154496}, \href
      {https://ui.adsabs.harvard.edu/abs/1971SoPh...16....3D} {16, 3}
    
    \bibitem[\protect\citeauthoryear{{Foukal} \& {Jokipii}}{{Foukal} \&
      {Jokipii}}{1975}]{Foukal1975}
    {Foukal} P.,  {Jokipii} J.~R.,  1975, \mn@doi [\apjl] {10.1086/181851}, \href
      {https://ui.adsabs.harvard.edu/abs/1975ApJ...199L..71F} {199, L71}
    
    \bibitem[\protect\citeauthoryear{{Gastine}, {Yadav}, {Morin}, {Reiners}  \&
      {Wicht}}{{Gastine} et~al.}{2014}]{gastine_2014}
    {Gastine} T.,  {Yadav} R.~K.,  {Morin} J.,  {Reiners} A.,   {Wicht} J.,  2014,
      \mn@doi [\mnras] {10.1093/mnrasl/slt162}, \href
      {https://ui.adsabs.harvard.edu/abs/2014MNRAS.438L..76G} {438, L76}
    
    \bibitem[\protect\citeauthoryear{{Gilman} \& {Foukal}}{{Gilman} \&
      {Foukal}}{1979}]{Gilman1979}
    {Gilman} P.~A.,  {Foukal} P.~V.,  1979, \mn@doi [\apj] {10.1086/157052}, \href
      {https://ui.adsabs.harvard.edu/abs/1979ApJ...229.1179G} {229, 1179}
    
    \bibitem[\protect\citeauthoryear{{Guerrero}, {Smolarkiewicz}, {Kosovichev}  \&
      {Mansour}}{{Guerrero} et~al.}{2013}]{guerrero_2013}
    {Guerrero} G.,  {Smolarkiewicz} P.~K.,  {Kosovichev} A.~G.,   {Mansour} N.~N.,
      2013, \mn@doi [\apj] {10.1088/0004-637X/779/2/176}, \href
      {https://ui.adsabs.harvard.edu/abs/2013ApJ...779..176G} {779, 176}
    
    \bibitem[\protect\citeauthoryear{{Hazra} \& {Choudhuri}}{{Hazra} \&
      {Choudhuri}}{2017}]{hazra_2017}
    {Hazra} G.,  {Choudhuri} A.~R.,  2017, \mn@doi [\mnras]
      {10.1093/mnras/stx2152}, \href
      {https://ui.adsabs.harvard.edu/abs/2017MNRAS.472.2728H} {472, 2728}
    
    \bibitem[\protect\citeauthoryear{{Hazra} \& {Choudhuri}}{{Hazra} \&
      {Choudhuri}}{2019}]{Hazra2019}
    {Hazra} G.,  {Choudhuri} A.~R.,  2019, \mn@doi [\apj]
      {10.3847/1538-4357/ab2718}, \href
      {https://ui.adsabs.harvard.edu/abs/2019ApJ...880..113H} {880, 113}
    
    \bibitem[\protect\citeauthoryear{{Hazra}, {Jiang}, {Karak}  \&
      {Kitchatinov}}{{Hazra} et~al.}{2019}]{Hazra2019b}
    {Hazra} G.,  {Jiang} J.,  {Karak} B.~B.,   {Kitchatinov} L.,  2019, \mn@doi
      [\apj] {10.3847/1538-4357/ab4128}, \href
      {https://ui.adsabs.harvard.edu/abs/2019ApJ...884...35H} {884, 35}
    
    \bibitem[\protect\citeauthoryear{{Hotta}, {Rempel}  \& {Yokoyama}}{{Hotta}
      et~al.}{2015}]{hotta_2015}
    {Hotta} H.,  {Rempel} M.,   {Yokoyama} T.,  2015, \mn@doi [\apj]
      {10.1088/0004-637X/798/1/51}, \href
      {https://ui.adsabs.harvard.edu/abs/2015ApJ...798...51H} {798, 51}
    
    \bibitem[\protect\citeauthoryear{{Howard} \& {Harvey}}{{Howard} \&
      {Harvey}}{1970}]{Howard_Harvey1970}
    {Howard} R.,  {Harvey} J.,  1970, \mn@doi [\solphys] {10.1007/BF02276562},
      \href {https://ui.adsabs.harvard.edu/abs/1970SoPh...12...23H} {12, 23}
    
    \bibitem[\protect\citeauthoryear{{Howe}}{{Howe}}{2009}]{howe_2009}
    {Howe} R.,  2009, \mn@doi [Living Reviews in Solar Physics]
      {10.12942/lrsp-2009-1}, \href
      {https://ui.adsabs.harvard.edu/abs/2009LRSP....6....1H} {6, 1}
    
    \bibitem[\protect\citeauthoryear{{Howe}, {Christensen-Dalsgaard}, {Hill},
      {Komm}, {Schou}  \& {Thompson}}{{Howe} et~al.}{2005}]{Howe2005}
    {Howe} R.,  {Christensen-Dalsgaard} J.,  {Hill} F.,  {Komm} R.,  {Schou} J.,
      {Thompson} M.~J.,  2005, \mn@doi [\apj] {10.1086/497107}, \href
      {https://ui.adsabs.harvard.edu/abs/2005ApJ...634.1405H} {634, 1405}
    
    \bibitem[\protect\citeauthoryear{{Karak}}{{Karak}}{2010}]{karak2010}
    {Karak} B.~B.,  2010, \mn@doi [\apj] {10.1088/0004-637X/724/2/1021}, \href
      {https://ui.adsabs.harvard.edu/abs/2010ApJ...724.1021K} {724, 1021}
    
    \bibitem[\protect\citeauthoryear{{Karak} \& {Choudhuri}}{{Karak} \&
      {Choudhuri}}{2011}]{karak2011}
    {Karak} B.~B.,  {Choudhuri} A.~R.,  2011, \mn@doi [\mnras]
      {10.1111/j.1365-2966.2010.17531.x}, \href
      {https://ui.adsabs.harvard.edu/abs/2011MNRAS.410.1503K} {410, 1503}
    
    \bibitem[\protect\citeauthoryear{{Karak}, {Jiang}, {Miesch}, {Charbonneau}  \&
      {Choudhuri}}{{Karak} et~al.}{2014a}]{karak_2014}
    {Karak} B.~B.,  {Jiang} J.,  {Miesch} M.~S.,  {Charbonneau} P.,   {Choudhuri}
      A.~R.,  2014a, \mn@doi [\ssr] {10.1007/s11214-014-0099-6}, \href
      {https://ui.adsabs.harvard.edu/abs/2014SSRv..186..561K} {186, 561}
    
    \bibitem[\protect\citeauthoryear{{Karak}, {Kitchatinov}  \&
      {Choudhuri}}{{Karak} et~al.}{2014b}]{karak2014b}
    {Karak} B.~B.,  {Kitchatinov} L.~L.,   {Choudhuri} A.~R.,  2014b, \mn@doi
      [\apj] {10.1088/0004-637X/791/1/59}, \href
      {https://ui.adsabs.harvard.edu/abs/2014ApJ...791...59K} {791, 59}
    
    \bibitem[\protect\citeauthoryear{{Kippenhahn} \& {Weigert}}{{Kippenhahn} \&
      {Weigert}}{1990}]{Kippenhahn1990}
    {Kippenhahn} R.,  {Weigert} A.,  1990, {Stellar Structure and Evolution}
    
    \bibitem[\protect\citeauthoryear{{Kitchatinov}}{{Kitchatinov}}{2013}]{kitchatinov_2013}
    {Kitchatinov} L.~L.,  2013, in {Kosovichev} A.~G.,  {de Gouveia Dal Pino} E.,
      {Yan} Y.,  eds,  IAU Symposium Vol. 294, Solar and Astrophysical Dynamos and
      Magnetic Activity. pp 399--410 (\mn@eprint {arXiv} {1210.7041}),
      \mn@doi{10.1017/S1743921313002834}
    
    \bibitem[\protect\citeauthoryear{{Kitchatinov} \& {Ruediger}}{{Kitchatinov} \&
      {Ruediger}}{1995}]{kitchatinov_1995}
    {Kitchatinov} L.~L.,  {Ruediger} G.,  1995, \aap, \href
      {https://ui.adsabs.harvard.edu/abs/1995A&A...299..446K} {299, 446}
    
    \bibitem[\protect\citeauthoryear{{Kuhn}, {Libbrecht}  \& {Dicke}}{{Kuhn}
      et~al.}{1988}]{kuhn_1988}
    {Kuhn} J.~R.,  {Libbrecht} K.~G.,   {Dicke} R.~H.,  1988, \mn@doi [Science]
      {10.1126/science.242.4880.908}, \href
      {https://ui.adsabs.harvard.edu/abs/1988Sci...242..908K} {242, 908}
    
    \bibitem[\protect\citeauthoryear{{Longcope} \& {Choudhuri}}{{Longcope} \&
      {Choudhuri}}{2002}]{Longcope_Choudhuri2002}
    {Longcope} D.,  {Choudhuri} A.~R.,  2002, \mn@doi [\solphys]
      {10.1023/A:1013896013842}, \href
      {https://ui.adsabs.harvard.edu/abs/2002SoPh..205...63L} {205, 63}
    
    \bibitem[\protect\citeauthoryear{{Matilsky}, {Hindman}  \& {Toomre}}{{Matilsky}
      et~al.}{2019}]{Matilsky2019}
    {Matilsky} L.~I.,  {Hindman} B.~W.,   {Toomre} J.,  2019, \mn@doi [\apj]
      {10.3847/1538-4357/aaf647}, \href
      {https://ui.adsabs.harvard.edu/abs/2019ApJ...871..217M} {871, 217}
    
    \bibitem[\protect\citeauthoryear{{Matilsky}, {Hindman}  \& {Toomre}}{{Matilsky}
      et~al.}{2020}]{Matilsky2020}
    {Matilsky} L.~I.,  {Hindman} B.~W.,   {Toomre} J.,  2020, \mn@doi [\apj]
      {10.3847/1538-4357/ab9ca0}, \href
      {https://ui.adsabs.harvard.edu/abs/2020ApJ...898..111M} {898, 111}
    
    \bibitem[\protect\citeauthoryear{{Moreno-Insertis}}{{Moreno-Insertis}}{1983}]{Moreno1983}
    {Moreno-Insertis} F.,  1983, \aap, \href
      {https://ui.adsabs.harvard.edu/abs/1983A&A...122..241M} {122, 241}
    
    \bibitem[\protect\citeauthoryear{{Rast}, {Ortiz}  \& {Meisner}}{{Rast}
      et~al.}{2008}]{rast_2008}
    {Rast} M.~P.,  {Ortiz} A.,   {Meisner} R.~W.,  2008, \mn@doi [\apj]
      {10.1086/524655}, \href
      {https://ui.adsabs.harvard.edu/abs/2008ApJ...673.1209R} {673, 1209}
    
    \bibitem[\protect\citeauthoryear{{Schou} et~al.,}{{Schou}
      et~al.}{1998}]{Schou1998}
    {Schou} J.,  et~al., 1998, \mn@doi [\apj] {10.1086/306146}, \href
      {https://ui.adsabs.harvard.edu/abs/1998ApJ...505..390S} {505, 390}
    
    \bibitem[\protect\citeauthoryear{{Spruit}}{{Spruit}}{1974}]{spruit_1974}
    {Spruit} H.~C.,  1974, \mn@doi [\solphys] {10.1007/BF00153665}, \href
      {https://ui.adsabs.harvard.edu/abs/1974SoPh...34..277S} {34, 277}
    
    \bibitem[\protect\citeauthoryear{{Vasil}, {Julien}  \& {Featherstone}}{{Vasil}
      et~al.}{2020}]{Vasil2020}
    {Vasil} G.~M.,  {Julien} K.,   {Featherstone} N.~A.,  2020, arXiv e-prints,
      \href {https://ui.adsabs.harvard.edu/abs/2020arXiv201015383V} {p.
      arXiv:2010.15383}
    
    \bibitem[\protect\citeauthoryear{{Wang}, {Sheeley}  \& {Nash}}{{Wang}
      et~al.}{1991}]{Wang1991}
    {Wang} Y.~M.,  {Sheeley} N.~R. J.,   {Nash} A.~G.,  1991, \mn@doi [\apj]
      {10.1086/170800}, \href
      {https://ui.adsabs.harvard.edu/abs/1991ApJ...383..431W} {383, 431}
    
    \makeatother
    \end{thebibliography}



\appendix
\section{Isochoric surfaces within the solar convection zone and temperature variations over them}
\label{appendix}

From the basic fluid dynamical equations, we arrive at 
Equation~\ref{eq1}
involving a differentiation of $S$ with respect to $\theta$ on a surface of constant $r$.  However, the rotation of the Sun makes the Sun slightly oblate so that the solar surface is not a surface of constant $r$.  The important question is how we connect a theory involving differentiation at constant $r$ to actual observations of the solar surface.

First of all, we argue that the oblateness of the Sun is extremely small.  We are not aware of any measurements of this oblateness.  However, one can make a rough estimate of this oblateness from theoretical considerations. If we take $\Omega$ to be constant inside the Sun for this approximate estimate, then we can introduce an effective potential
\begin{equation}
 \Phi_{\rm eff} = - \frac{G M_{\odot}}{r} - \frac{1}{2} | {\bf \Omega} \times {\bf r} |^2
 \label{ae1}
\end{equation}
inside the convection zone, which has to be constant over isobaric surfaces (see, for example, \citet{Choudhuri1998book}, Section~9.3). If $\Delta r$ is the extension of such a surface in the equatorial 
region compared to the polar region, then we can equate $\Phi_{\rm eff}$ in the polar and equatorial regions to obtain

\begin{equation}
 - \frac{G M_{\odot}}{r} = - \frac{G M_{\odot}}{r + \Delta r} - \frac{1}{2} \Omega^2 r^2,\nonumber
\end{equation}
from which
\begin{equation}
\Delta r = \frac{1}{2} \frac{\Omega^2 r^4}{G M_{\odot}}.
\label{ae2}
\end{equation}
Since the mass contained within the convection zone is small compared to the solar mass, we can
take $M_{\odot}$ to be the total solar mass for estimating the oblateness of isobaric surfaces
within the convection zone.  Using $\Omega/2 \pi = 420$ nHz and $r = 0.85$\Rsun, we find
\begin{equation}
\Delta r = 3.3 \; {\rm km},
\label{ae3}
\end{equation}
which is less than 0.0005\% of the solar radius. The expected oblateness of the Sun is minuscule.

We now focus our attention on isochoric surfaces over which the entropy differential would
be related with the temperature differential by Equation~\ref{eq2}.
We denote the distance measured along an isochore by $l$. It is pointed out in Appendix A of \citet{Choudhuri2021a} that we approximately have

\begin{equation}
\frac{dS}{dl} = \left( \frac{\pa S}{\pa \theta}\right)_r \frac{d \theta}{d l}. 
\label{ae4}
\end{equation}
Using this relation, we can put Equation~\ref{eq1} in the form
\begin{equation}
r \sin \theta \frac{\pa}{\pa z} \Omega^2 = \frac{g}{\gamma C_V} \frac{d S}{d l}, 
\label{ae5}
\end{equation}
where $d S/ dl$ is the derivative of entropy $S$ along the isochore and we have used $d l = r d \theta$
in view of the very small oblateness of the Sun.  
We can now substitute Equation~\ref{eq2} in Equation~\ref{ae5} to obtain

\begin{equation}
r \sin \theta \frac{\pa}{\pa z} \Omega^2 = \frac{g}{\gamma T} \frac{d}{d l} \Delta T.
\label{ae6}
\end{equation}
Now, as we move along an isochore, the change in $l$ 
is accompanied by a change in $\theta$.  We have
\begin{equation}
\frac{d}{d l} \Delta T = \frac{1}{r}
\left(\frac{d}{d \theta} \Delta T \right)_{\rm isochore}.
\label{ae7}
\end{equation}
Note that here the differenetiation is with respect to
the changing value of $\theta$ along an isochore.
On substituting Equation~\ref{ae7} into Equation~\ref{ae6}, we are readily led to Equation~\ref{eq3}. 

One fact is clear from Equation~\ref{eq3}.  If the Sun did not have a thermal wind (which would require $\Omega$
not to vary along $z$ for consistency), then the temperature would be constant over the
isochoric surface.  The isochoric surface would then be a isothermal surface and hence also an
isobaric surface.  Now,
the observed solar surface is a surface at which the optical depth becomes 1, the opacity
depending on both density and temperature. In the absence of the thermal wind, the isochoric, isothermal
and isobaric surfaces would coincide and the observed surface of the Sun would be one such surface.
To obtain the temperature variation on the solar surface, we need to find the temperature difference
on isochoric surfaces.  For this purpose, we can use Equation~\ref{eq3}.


If $\Omega (r, \theta)$ is given, then
Equation~\ref{eq3} can be integrated to obtain
$\Delta T$ along an isochore.  Because of the miniscule oblateness of the isochore, we can regard the isochore
to be a spherical surface.  However, at a conceptual level, we shall keep in mind that we would be considering
$\Delta T$ on isochoirc surfaces, since this will eventually enable us to compare with observational data
on the solar surface.

We point out one other thing, which can sometimes be
a source of confusion. By the chain rule of 
differentiation, we have
\begin{equation}
\frac{d}{dl}\Delta T = \left( \frac{\partial T}{\partial r} \right)_{\theta} \frac{d r}{dl}
+ \left( \frac{\partial T}{\partial \theta} \right)_r \frac{d \theta}{dl}. 
\label{ae8}
\end{equation}
Even though $d r/ dl$ is very small, the first term on
the right hand side of Equation~\ref{ae8} cannot be
neglected with respect to the second term because $(\partial T/ \partial r)_{\theta}$ is much larger
than $(1/r)(\partial T / \partial \theta)_r$.  As a result, we cannot write down an equation for temperature
similar to Equation~\ref{ae4} for entropy. However, when
we connect the theory with observations, the observed
temperature variations of the solar surface have to be
compared with $((d/d \theta) \Delta T)_{\rm isochore}$
appearing in Equation~\ref{eq3} and not with $(\partial T / \partial \theta)_r$ which is not of much interest to us.



\bsp	
\label{lastpage}
\end{document}